\documentclass[11pt, a4paper]{article}

\usepackage{amssymb,amsmath,graphicx,color,multirow}
\begin{document}

\vspace*{-1.8cm}
\begin{flushright}
{\bf\normalsize LAL 11-245}\\
%\vspace*{-0.5cm}
{\normalsize October 2011}
\end{flushright}
\vspace*{2cm}
\begin{center}
{\huge\bf Physics Opportunities at the Next Generation \\
\vspace*{0,2cm} of Precision Flavor Physics}\\
\vspace*{1.5 cm}
{\Large\bf Marco Ciuchini}\\
\vspace*{0.5cm}
{\large\it INFN Sezione di Roma Tre,\\Via della Vasca Navale 84, I-00146 Rome, Italy}\\
\vspace*{1.5 cm}
{\Large\bf Achille Stocchi}\\
\vspace*{0.5cm}
{\large\it  LAL, Univ Paris-Sud, CNRS/IN2P3, Orsay, France}
\end{center}
\vspace*{2 cm}

\begin{abstract}
Starting with next-generation experiments, flavor physics fully enters the
era of precision measurements. The focus shifts from testing the Standard Model to
finding and characterizing new physics contributions. We review the opportunities
offered by future flavor experiments, discussing the expected sensitivities of the most
important measurements. We also present some examples of measurable deviations from
the Standard Model in the flavor sector generated in a selection of
new physics models, demonstrating the major contribution that precision flavor physics
could give to the effort of going beyond the Standard Model.
\end{abstract}
\vspace*{1cm}

\section{Introduction}
\label{sec:intro}
Up to now the Standard Model (SM) managed to pass all the experimental challenges unscathed,
providing an overall good description of particle physics up to the energy scales probed
in experiments so far, namely hundreds of GeV.
The Higgs boson, the last missing building block of the SM, is being searched for
at the LHC experiments ATLAS and CMS, which will be able to find or exclude it.

In spite of the phenomenological success, however, the SM is not satisfactory for several
theoretical reasons, including the instability of the fundamental scale of weak interactions,
the Fermi scale, against radiative corrections, the absence of a dark matter candidate, an amount
of CP violation too small to account for the matter-antimatter asymmetry in the universe,
the lack of an explanation for the origin of flavor and CP violation and the non-unified
description of the interactions, with gravity completely missing.

For these reasons, the SM is regarded as a low-energy theory bound to fail at some
energy scale larger than the Fermi scale, where New Physics (NP) effects become important.
The search for these effects beyond the SM is the main goal of particle physics in the next
decades, both at present and future experimental facilities.

The most straightforward way to search for NP is producing and observing new
particles in colliders. To this end, the key ingredient is the available center-of-mass
energy: the higher the energy, the heavier the particles one can produce, thus probing
higher NP scales. Pushing forward the energy frontier then means building colliders with
higher and higher center-of-mass energy. The problem with direct searches is that their
success depends on the unknown production thresholds of the
new particles. Some of the SM problems, however, have solutions, such as the natural stabilization
of the Fermi scale or the WIMPs (Weakly Interacting Massive Particles) as dark matter
candidates, which point to NP at the TeV scale giving to the LHC good chances to discover it.

A complementary way to reveal NP is measuring the effect of virtual heavy particle
exchange in processes involving only SM particles as external states. 
For this kind of searches, called indirect, the production threshold is not
an issue. Since quantum effects become typically smaller as the mass of the virtual
particles increases, higher NP scales are probed by increasing the precision of
the measurements, controlling at same time the SM contributions with an accuracy
sufficient to disentangle genuine NP effects from SM uncertainties. Progress at the
intensity frontier entails building experimental facilities which can deliver
high-intensity beams at energies where one can make a set of precision measurements 
not limited by systematic and theoretical uncertainties.

Flavor physics is the best candidate as a tool for indirect NP searches for
several reasons. Flavor Changing Neutral Currents (FCNC), neutral
meson-antimeson mixing and CP violation occur at the loop level in the SM and
therefore are potentially subject to large virtual corrections
from new heavy particles. In addition, flavor violation in the SM is governed
by weak interactions and, in the case of quarks, suppressed by small mixing angles.
These features are not necessarily shared by NP which could then produce
very large effects.

On quite general grounds, the indirect NP search in flavor processes explores a
parameter space including the NP scale and the NP flavor- and CP-violating
couplings. In specific models, these are related to fundamental parameters such
as masses and couplings of new particles. In particular, an observable NP effect
could be generated by small NP scales and/or large couplings.
Indeed the inclusion in the SM of generic NP flavor-violating terms
with $\mathcal{O}(1)$ couplings is known to violate present experimental
constraints unless the NP scale is pushed to large values, from $10^2$ to $10^4$ TeV
depending on the flavor sector, with the larger scales required by kaon
physics~\cite{Bona:2007vi}
%~\cite{Bona:2007vi,Isidori:2010kg}.

The difference between the NP scale emerging from flavor physics when flavor couplings
are not suppressed and the TeV scale suggested by the solution of the SM problems
mentioned above is usually referred to as the NP flavor problem: NP models with
new particles at the TeV scale need a mechanism to suppress NP contributions
to FCNC and CP-violating processes.
On the other hand, this clearly indicates that flavor physics has either the potential
to push the explored NP scales well beyond the TeV region or, if the NP scale is indeed
close to $1$ TeV, that the NP flavor structure is non-trivial, calling for the experimental
determination of new flavor-violating couplings in order to elucidate the suppression
mechanism at work.

In spite of its potential, so far flavor physics has not provided a clear NP signal,
although deviations with low significance are present here and there.
This failure, however, should not be considered discouraging. Broadly speaking,
flavor physics probed and excluded only the region of small masses/large NP flavor
couplings. On the other hand, we should not forget that flavor physics already proved
itself effective in predicting the existence and in some cases the mass of most of the
heavy particles which are nowadays part of the SM. The list of these successes includes
the prediction of the existence of the charm quark from the suppression of
the rate $K^0\to\mu^+\mu^-$ in the early seventies~\cite{Glashow:1970gm},
the prediction of the existence of the third generation
from the measurement of CP violation in $K$--$\bar K$ mixing, again in the
seventies~\cite{Kobayashi:1973fv} and the indication of a heavy mass for the top quark
from the measurements of the semileptonic $B$ decay rates and the $B$--$\bar B$ mass difference in
the late eighties~\cite{Bigi:1988jr}.
%~\cite{Bigi:1988jr,Lusignoli:1991bm}.
%In addition, a more precise value of the top mass was anticipated by the
%fit to LEP electroweak precision data just before the discovery at the Tevatron
%in 1995. While definitely not related to flavor physics, this is still a success of indirect
%searches through loop effects using precise measurements.
The lesson to be learned from this short historical excursus is that flavor
physics can indeed provide very valuable information on unknown heavy particles
and it is therefore worth pushing experiments and theory to higher precision in
order to probe heavier particles and/or smaller flavor couplings.

In this paper, we give an overview of the physics opportunities opened by the
next-generation flavor experiments, mainly the two approved super $B$-factories SuperB
in Italy and Belle-II/SuperKEKB in Japan. These are actually super flavor-factories,
expected to produce precision measurements in $B$, $D$ and $\tau$ physics, increasing by
two orders of magnitude the luminosity of their predecessors PEP-II and KEKB.
The two super-flavor factories are planned to start to take data in the second part of
this decade and to collect more than 50 ab$^{-1}$ integrated luminosity by around 2020.
It has to be noted that by $\sim 2015$ the LHCb experiment at LHC will collect the design
integrated luminosity of 10 fb$^{-1}$ and an upgrade (after 2020) is actually under study
to considerably increase the collected data sample up to 50 fb$^{-1}$.

We begin with a brief primer on flavor and CP violation in the SM, introducing also the 
some common flavor observables, in section~\ref{sec:primer}. We then summarize 
in section~\ref{sec:status} the present knowledge of the SM flavor sector and point out
some tensions which could be the harbingers of physics beyond the SM. 
The most interesting measurements in the physics programme of the next-generation
flavor factories are reviewed in section~\ref{sec:opportunities}, giving estimates of
the expected statistical and systematic errors. The selection of these measurements is
done taking into account the physics program which should have been already accomplished 
by LHCb. In section~\ref{sec:bsm}, we discuss the impact of these measurements on a
selection of NP  models in different scenarios, giving few examples of the potential
implications of precision  flavor physics.
Finally, our conclusions are presented in section~\ref{sec:conclusions}.

We are particularly indebted with the authors of
refs.~\cite{Bona:2007qt,Browder:2007gg,Hitlin:2008gf,Aushev:2010bq,O'Leary:2010af},
as many arguments and results presented in this review are based on their studies.

\section{Flavor and CP violation: a short primer}
\label{sec:primer}
% The origin of flavor and CP violation still lacks a theoretical explanation.
% The SM only offers a phenomenological model of flavor including all Yukawa interactions
% compatible with its gauge symmetry and particle content.
In this section we briefly review the SM flavor structure,
focusing mainly on the quark sector and collecting some basic formulae
used throughout the review.

\subsection{Flavor and CP violation in the Standard Model}
The SM matter content is organized as follows:
\begin{eqnarray}
& &\mathrm{quarks: }\quad Q_L^i\, \left(3,2,\frac{1}{6}\right),\quad U_R^i\,
\left(3,1,\frac{2}{3}\right),\quad D_R^i\,\left(3,1,-\frac{1}{3}\right)\nonumber\\
& &\mathrm{leptons: }\quad L_L^i\, \left(1,2,-\frac{1}{2}\right),\quad E_R^i\,
\left(1,1,-1\right)\,,
\end{eqnarray}
where the transformation properties under the gauge groups
$SU(3)_C\times SU(2)_L\times U(1)_Y$ are indicated.
%by the representation for non-abelian
%groups and by the charge for the abelian one.
The index $i$ spans the flavor space. In the SM with three generations, one has
\begin{eqnarray}
&&Q_L^i=\begin{pmatrix}u_L^i\\ d_L^i\end{pmatrix}\,,\quad
L_L^i=\begin{pmatrix}\nu^i_L,\\ \ell^i_L\end{pmatrix}\,,\quad
U_R^i=u_R^i\,,\quad D_R^i=d_R^i\,,\quad E_R^i=\ell^i_R\,,\nonumber\\
&& u^i=(u,c,t)\,,\quad d^i=(d,s,b)\,,\quad \ell=(e,\mu^-,\tau^-)\,,\quad
\nu^i=(\nu^e,\nu^\mu,\nu^\tau)\,.
\end{eqnarray}
The subscripts $L$ and $R$ denote the left-handed and right-handed components of the fields.
The SM Lagrangian can be written as
\begin{equation}
\mathcal{L}^\mathrm{SM}=\mathcal{L}^\mathrm{Yang-Mills}+\mathcal{L}^\mathrm{Dirac+gauge}+
\mathcal{L}^\mathrm{Higgs}+\mathcal{L}^\mathrm{Yukawa}\,.
\end{equation}
In the absence of $\mathcal{L}^\mathrm{Yukawa}$, the SM Lagrangian is invariant under
the global symmetry 
\begin{equation}
 G_\mathrm{flavor}=U(3)_Q \otimes U(3)_U \otimes U(3)_D \otimes U(3)_L \otimes U(3)_E\,,
\label{eq:gflavor}
\end{equation}
corresponding to the separate unitary rotation of each matter multiplet in flavor space.
$\mathcal{L}^\mathrm{Yukawa}$ contains the Yukawa interactions between the Higgs doublet
and the matter fields:
\begin{equation}
 \mathcal{L}^\mathrm{Yukawa}=\bar Q_L^i Y_u^{ij} U_R^j \widetilde H + \bar Q_L^i Y_d^{ij}
D_R^j H + \bar L_L^i Y_\ell^{ij} E_R^j H +\mathrm{H.c.}\,,
\label{eq:Y}
\end{equation}
where $H$ is the Higgs doublet transforming as $\left(1,2,1/2\right)$ and
$\widetilde H=i\sigma_2 H^*\,\left(1,2,-1/2\right)$. The Yukawa couplings $Y_{u,d,\ell}$ 
are $3\times 3$ complex matrices in flavor space (the sum over the
flavor indices $i,j$ is understood).
The presence of non-vanishing Yukawa couplings breaks this large group down to few
$U(1)$ factors,
\begin{equation}
 G_\mathrm{flavor} \xrightarrow[Y_{u,d,\ell}\neq 0]{} U(1)_B \otimes U(1)_e \otimes U(1)_\mu
\otimes U(1)_\tau\,,
\end{equation}
which account for the conservation of the barion and the three lepton numbers.

% The structure of the Yukawa matrices dictates the SM flavor properties.
% These three generic complex matrices contain $3\times 18=54$ free parameters. However many
% of them are not physical and can be removed by rotating the fermion fields. In the
% quark sector, for instance, only $10$ parameters out of $36$ are observable:
% $9$ moduli and $1$ phase~\cite{hep-ph/9911321}. The Yukawa matrices
They can be diagonalized
using the singular value decomposition, giving
\begin{equation}
 Y_u^\mathrm{diag}=U_L^u Y_u U_R^{u\dagger}\,,\quad  Y_u^\mathrm{diag}=U_L^d Y_u
U_R^{d\dagger}\,,
\label{eq:svd}
\end{equation}
where $U_{L,R}^{u,d}$ are four unitary matrices and $Y_{u,d}^\mathrm{diag}$ are diagonal
matrices with real non-negative matrix elements. Therefore it is always possible to choose a
flavor basis for the quark fields where, for example,
\begin{equation}
 Y_d=Y_d^\mathrm{diag}\,,\quad  Y_u=V^\dagger Y_u^\mathrm{diag}\,. 
\end{equation}
$V=U_L^u U_L^{d^\dagger}$ is the Cabibbo-Kobayashi-Maskawa
matrix~\cite{Cabibbo:1963yz,Kobayashi:1973fv}
% Clearly $Y_d^\mathrm{diag}$ and $Y_u^\mathrm{diag}$ contain $6$ real parameters
% while $V_{CKM}$ can be parameterized using the remaining $3$ angles and $1$ phase.
which contains all flavor non-diagonal couplings and the only CP-violating phase
of the SM Lagrangian (notice however that the flavor symmetry is broken also by
$Y_{u,d}^\mathrm{diag}$ as long as the diagonal entries are different).
Clearly $V$ originates from the misalignment
in flavor space of the components of the left-handed quark doublet. 

After the Electroweak Symmetry Breaking (EWSB), the Yukawa interactions generate the
fermion masses.
% The Higgs doublet acquires a vacuum expectation value $v$.
% becoming
% \begin{eqnarray}
%  H \xrightarrow[EWSB]{}\begin{pmatrix}0\\\frac{v}{\sqrt{2}}+h\end{pmatrix}\,,
% \end{eqnarray}
% where $h$ is the physical Higgs field.
%The mass terms are then obtained from eq.~(\ref{eq:Y}) replacing $H\to (0,v/\sqrt{2})$,
%so that
Replacing $H\to (0,v/\sqrt{2})$ in eq.~(\ref{eq:Y}), one obtains
\begin{equation}
\mathcal{L}^\mathrm{mass}=\bar u_L^i M^u_{ij} u_R^j+\bar d_L^i M^d_{ij} d_R^j+
\mathrm{H.c.}\,,\quad
 M^u = Y_u \frac{v}{\sqrt{2}},\quad M^d = Y_d \frac{v}{\sqrt{2}}\,.
\end{equation}
One can go to the mass eigenstate basis, where $M^u$ and $M^d$ are both diagonal,
applying unitary flavor transformations to the quark fields, namely making the
replacement
\begin{equation}
 u_{L,R}\to U_{L,R}^{u\dagger} u_{L,R}\,,\quad d_{L,R}\to U_{L,R}^{d\dagger} d_{L,R}\,,
\end{equation}
where $U_{L,R}^{u,d}$ are the four unitary matrices defined in eq.~(\ref{eq:svd}).
The only leftover of these transformations in the SM Lagrangian is the CKM matrix
appearing in the charged current weak interactions
\begin{equation}
\mathcal{L}^{cc}=\frac{g_2}{\sqrt{2}}\bar u^i_L \gamma_\mu
V_{ij}  d^j_L\, W^\mu+\mathrm{H.c.}\,
\end{equation}
where $g_2$ is the $SU(2)$ coupling constant.

Let us have a closer look at the CKM matrix $V$.
A generic $3\times 3$ unitary matrix can be represented
using $3$ Euler angles and $6$ phases, but $5$ of the latter can be removed by re-phasing
the quark fields. Therefore $V$ can be parameterized by $3$ angles
$\theta_{12},\theta_{23},\theta_{13}\in [0,\pi/2]$ and $1$ phase $\delta\in [-\pi,\pi)$.
Notice that the choice of the paramters is not unique.
In the parameterization used by the PDG~\cite{PDG}, the CKM matrix reads
\begin{equation}
V\equiv\left(\begin{smallmatrix}
V_{ud} & V_{us} & V_{ub}\\
V_{cd} & V_{cs} & V_{cb}\\
V_{td} & V_{ts} & V_{tb}
\end{smallmatrix}\right)
=\left(\begin{smallmatrix}
c_{12}c_{13}&  s_{12}c_{13}& s_{13}\, e^{-i\delta}\\
-s_{12}c_{23}-c_{12}s_{13}s_{23}\, e^{i\delta} & c_{12}c_{23}-s_{12}s_{13}s_{23}\,e^{i\delta} & c_{13}s_{23}\\
s_{12}s_{23}-c_{12}s_{13}c_{23}\, e^{i\delta} & -c_{12}s_{23}-s_{12}s_{13}c_{23}\, e^{i\delta} & c_{13}c_{23}
\end{smallmatrix}\right),
\label{eq:V}
\end{equation}
where $s_{ij}=\sin\theta_{ij}$ and $c_{ij}=\cos\theta_{ij}$.

The CP-violating phase $\delta$ is not a physical parameter as it depends on
the phase conventions of the quark fields. An invariant condition for CP violation
in the SM is given by ~\cite{Jarlskog:1985ht}
\begin{eqnarray}
&&0\neq\det\left[M^u M^{u\dagger},M^d M^{d\dagger}\right]=\nonumber\\
&&\quad 2iJ\,(m_u^2-m_c^2)(m_c^2-m_t^2)(m_t^2-m_u^2)(m_d^2-m_s^2)(m_s^2-m_b^2)(m_b^2-m_d^2)\,,~~~~
\end{eqnarray}
where the Jarlskog invariant
%\begin{equation}
$J=\vert\mathrm{Im}(V_{ij}V_{kl}V_{il}^*V_{kj}^*)\vert$\ (for all $i\neq k$ and $j\neq l$)
%\end{equation}
is independent of the convention chosen for the CKM matrix. In terms of the parameters
used in eq.~(\ref{eq:V}), $J=s_{12}s_{13}s_{23}c_{12}c_{13}^2c_{23}\sin\delta$.
Therefore CP violation in the SM requires $\delta\neq 0,\pi$, non-degenerate masses in the
up and down sectors and non-trivial mixing angles $\theta_{ij}\neq 0,\pi/2$. \\
\vspace*{0.5cm}

\begin{figure}[h]
\centerline{\includegraphics{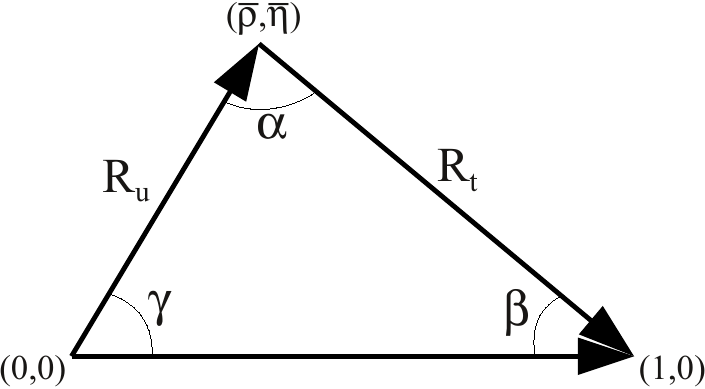}}
\caption{The Unitarity Triangle}
\label{fig:UT}
\end{figure}

The unitarity of $V$ implies $9$ conditions involving products of rows
(or columns) of the form $\sum_{k} V_{ik}V^*_{jk}=\delta_{ij}$.
The off-diagonal relations are called triangular as they define triangles in the complex
plane. Remarkably, the area of all these triangles is a constant equal to $J/2$ and thus a
measure of CP violation in the SM. One of them, of particular
phenomenological interest, is referred to as the Unitarity Triangle (UT):
\begin{equation}
V_{ud}V_{ub}^*+V_{cd}V_{cb}^*+ V_{td}V_{tb}^*=0\,.
\end{equation}
The UT can be rewritten in the normalized form 
\begin{equation}
R_t\,e^{-i\beta}+R_u\,e^{i\gamma}=1\,,
\label{eq:UT}
\end{equation}
with
\begin{equation}
R_t=\left|\frac{V_{td}V_{tb}^*}{V_{cd}V_{cb}^*}\right|,
R_u=\left|\frac{V_{ud}V_{ub}^*}{V_{cd}V_{cb}^*}\right|,
\beta=\arg\left(-\frac{V_{cd}V_{cb}^*}{V_{td}V_{tb}^*}\right),
\gamma=\arg\left(-\frac{V_{ud}V_{ub}^*}{V_{cd}V_{cb}^*}\right),
\end{equation}
being two sides and two angles as sketched in fig.~\ref{fig:UT}.
The third side is the unity vector, while the third angle is
$\alpha=\pi-\beta-\gamma=\arg(-V_{td}V_{tb}^*/(V_{ud}V_{ub}^*))$.
Similarly, it is useful to define the angle
$\beta_s=\arg(-V_{ts}V_{tb}^*/(V_{cs}V_{cb}^*))$ from the triangle
$V_{us}V_{ub}^*+V_{cs}V_{cb}^*+ V_{ts}V_{tb}^*=0$,
as the phase of the $B_s$--$\bar B_s$ mixing amplitude is $-2\beta_s$ in the phase convention
of eq.~(\ref{eq:V}). The UT sides and angles are observables and we discuss their present
determination from $K$- and $B$-meson physics in section~\ref{sec:status}.

Given the definition in eq.~(\ref{eq:UT}), all the information related to the UT
is encoded in one complex number
\begin{equation}
 \bar\rho+i\,\bar\eta=R_u\, e^{i\gamma}
\end{equation}
corresponding to the coordinates $(\bar\rho,\bar\eta)$ in the complex plane
of the only non-trivial apex of the UT.

It is worth mentioning another popular CKM parameterization introduced by
Wolfenstein~\cite{Wolfenstein:1983yz} which allows to write an expansion of
the CKM matrix in terms of a small parameter $\lambda$:
\begin{equation}
V=\left(\begin{smallmatrix}
 1-\frac{\lambda^2}{2} & \lambda & A\lambda^3(\rho-i\,\eta)\\
 -\lambda &  1-\frac{\lambda^2}{2} & A\lambda^2\\
 A\lambda^3(1-\rho-i\,\eta) & -A\lambda^2 & 1
\end{smallmatrix}\right)+\mathcal{O}(\lambda^4)
\end{equation}
This parameterization makes explicit the hierarchy of CKM matrix elements,
showing that quark flavor-changing transitions are suppressed in the SM.
The exact relations between the PDG and the Wolfenstein parameterizations
are
\begin{equation}
\lambda=\sin\theta_{12},\quad A=\frac{\sin\theta_{23}}{\sin^2\theta_{12}},\quad
\rho=\frac{\sin\theta_{13}\cos\delta}{\sin\theta_{12}\sin\theta_{23}},\quad
\eta=\frac{\sin\theta_{13}\sin\delta}{\sin\theta_{12}\sin\theta_{23}}\,.
\end{equation}
Notice that $\lambda\sim 0.22$ (the sine of the Cabibbo angle $\theta_{12}$) is indeed a good
expansion parameter. The relation between the UT apex coordinates and the Wolfenstein
parameters is
\begin{equation}
 \rho+i\,\eta=\sqrt{\frac{1-A^2\lambda^4}{1-\lambda^2}}
 \frac{\bar\rho+i\,\bar\eta}{1-A^2\lambda^4(\bar\rho+i\,\bar\eta)}
 \simeq \left(1+\frac{\lambda^2}{2}\right)\left(\bar\rho+i\,\bar\eta\right)+
 \mathcal{O}(\lambda^4)\,.
\end{equation}
showing that $\bar\rho=\rho$ and $\bar\eta=\eta$  at the lowest order in $\lambda$. 

For massless neutrinos, the SM lepton sector is flavor diagonal and CP conserving,
as $Y_\ell$ can be chosen real and diagonal.
Once the SM is trivially extended to include right-handed neutrinos and account for
neutrino Dirac masses, introducing an additional Yukawa matrix $Y_\nu$, the formalism for
lepton flavor and CP violation is similar to the quark one. The lepton mixing
matrix, called the Pontecorvo-Maki-Nakagawa-Sakata matrix,~\cite{Pontecorvo:1957cp}, is
parameterized much as the CKM matrix. The only difference is that the possibility of having
Majorana mass terms for neutrinos allows for two additional CP-violating phases. In spite of
these formal similarities, the flavor phenomenology of the lepton sector, in particular of
neutrinos, is rather different from the quark one and its discussion goes beyond the scope
of this primer. We just remind that, given the smallness of neutrino masses,
charged lepton flavor violation is negligible in the SM.

% In summary, the main flavor properties of the SM are:
% \begin{itemize}
%  \item[-] flavour-changing transitions are mediated by weak interactions only. The
% the strenght of the transitions is dictated by the relevant CKM matrix elements;
%  \item[-] flavour-changing neutral currents are absent at the tree level;
%  \item[-] the CKM matrix is hierarchical and can be expanded in $\lambda$
% (the sine of the Cabibbo angle);
%  \item[-] there is only one source of CP violation in the CKM matrix, correlating all
% CP-violating phenomena;
%  \item[-] the strength of CP violation is suppressed by mixing angles. Small CP violation
% does not imply a small phase in the CKM matrix;
%  \item[-] even if neutrino masses are taken into account in the SM, charged lepton flavor
% violation is negligibly small.
% \end{itemize}

\subsection{Meson mixing and CP violation}

Most of flavor physics deals with flavor-changing transitions involving mesons.
As far as decays are concerned, CP violation
appears as a difference between the rates of a decay and its CP-conjugate,
accounted for by the direct CP asymmetry
\begin{equation}
 A_\mathrm{CP}=\frac{\Gamma(\bar M\to \bar f)-\Gamma(M\to f)}
{\Gamma(\bar M\to\bar f)+\Gamma(M\to f)}=\frac{\vert\bar A\vert^2-\vert A \vert^2}
{\vert\bar A\vert^2+\vert A \vert^2}\,,
\label{eq:acpdir}
\end{equation}
where $M$ is the decaying meson, $f$ is the final state, $\bar M$ and $\bar f$ are
the CP-conjugate states and $A$ and $\bar A$ are the two decay amplitudes.
Charged mesons can only violate CP in the decay.
On the other hand, neutral mesons, with the exception of the pion,
are subject to the mixing phenomenon, {\em i.e.}\ the mass eigenstates are
a superposition of the flavor ones. In this case, CP violation can also occur
in the mixing itself and in the interference between mixing and decay, giving
additional opportunities to observe it. We briefly summarize in the following
the main formulae related to meson mixing and CP violation.

The time evolution of a system of unstable neutral meson-antimeson states
$M$--$\bar M$ can be described by a $2\times 2$ non-Hermitean matrix Hamiltonian
$\hat H$
\begin{equation}
 i \frac{d}{dt}\begin{pmatrix} |M(t)\rangle \\ |\bar M(t) \rangle
\end{pmatrix}=\hat H \begin{pmatrix} |M(t)\rangle\\ |\bar M(t)\rangle
\end{pmatrix}=\left(\hat m-\frac{i}{2}\hat \Gamma\right)
\begin{pmatrix} |M(t)\rangle\\ |\bar M(t)\rangle \end{pmatrix}\,.
\end{equation}
$\hat H$ can be decomposed using the two Hermitean matrices $\hat m$ and $\hat\Gamma$
representing its dispersive and absorbitive part respectively. In particular,
assuming CPT invariance, one can write
\begin{equation}
 \hat m=\begin{pmatrix}m & m_{12}\\ m_{12}^* & m\end{pmatrix}\,,\quad
 \hat \Gamma=\begin{pmatrix}\Gamma & \Gamma_{12}\\ \Gamma_{12}^* & \Gamma\end{pmatrix}\,.
\end{equation}
The eigenstates of $\hat H$, denoted as $|M_{L,H}\rangle$, can be written as
\begin{equation}
|M_{L,H}\rangle=\frac{1}{\sqrt{1+|q/p|^2}}\left(|M\rangle \pm q/p\, |\bar M
 \rangle\right)\,,\quad
 q/p=-\sqrt{\frac{m_{12}^*-\frac{i}{2}\Gamma_{12}^*}
      {m_{12}-\frac{i}{2}\Gamma_{12}}}\,,
\label{eq:qop}
\end{equation}
where $L$ ($H$) corresponds to the upper (lower) sign.
The corresponding eigenvalues are
\begin{equation}
m_{L,H}-\frac{i}{2}\Gamma_{L,H}=m-\frac{i}{2}\Gamma \mp
\sqrt{\left(m_{12}-\frac{i}{2}\Gamma_{12}\right)\left(m_{12}^*-
\frac{i}{2}\Gamma_{12}^*\right)}\,,
\end{equation}
giving the mass and width differences between the two eigenstates
\begin{eqnarray}
&&\Delta m=m_H-m_L=2\,\mathrm{Re}\sqrt{\left(m_{12}-\frac{i}{2}\Gamma\right)
  \left(m_{12}^*-\frac{i}{2}\Gamma^*\right)}\,,\nonumber\\
&&\Delta \Gamma=\Gamma_H-\Gamma_L=-4\,\mathrm{Im}\sqrt{\left(m_{12}-
 \frac{i}{2}\Gamma\right)\left(m_{12}^*-\frac{i}{2}\Gamma^*\right)}\,.
\label{eq:dmdg}
\end{eqnarray}
These observables, or the corresponding dimensionless variables
$x=\Delta m/\Gamma$ and $y=\Delta\Gamma/2\Gamma$ ($\Gamma$ is the average
lifetime), characterize the mixing.
In addition, the absolute value $\vert q/p\vert$ is also observable. The deviation
of $\vert q/p\vert$ from one is a measure of CP violation in the mixing,
as shown by the expression of the semileptonic CP asymmetry~\footnote{This formula
is correct for $M=K^0,\,B_d,\,B_s$. For $D^0$ mesons, the leptons in the final state
are CP conjugated.}
\begin{eqnarray}
 A_\mathrm{SL}&=&\frac{\Gamma(\bar M\to X\ell^+\nu_\ell)-\Gamma(M\to X\ell^-\bar \nu_\ell)}
{\Gamma(\bar M\to X\ell^+\nu_\ell)-\Gamma(M\to X\ell^-\bar \nu_\ell)}\nonumber\\
&=&\frac{1-\vert q/p\vert^4}{1+\vert q/p\vert^4}=\frac{\mathrm{Im}(m_{12}^*\Gamma_{12})}{|m_{12}|^2+
\frac{1}{4}|\Gamma_{12}|^2}\,.
\end{eqnarray}
Depending on the meson, these formulae can be approximated in different
ways. For $B$ mesons, where $\Gamma_{12}\ll m_{12}$, one finds
\begin{eqnarray}
&&\Delta m\simeq 2\vert m_{12}\vert\,,\quad \Delta \Gamma\simeq \Delta m\,
  \mathrm{Re} \left(\frac{\Gamma_{12}}{m_{12}} \right)\,,\nonumber\\
&& \vert q/p\vert \simeq 1-\frac{1}{2}\mathrm{Im}
\frac{\Gamma_{12}}{m_{12}}\,,\quad A_{SL}\simeq \mathrm{Im} \left(\frac{\Gamma_{12}}{m_{12}} \right)\,.
\end{eqnarray}
For the $D$ meson such simplification is not possible and the full expressions
have to be used. Besides, eqs.~(\ref{eq:dmdg}) only give
the short-distance part of the corresponding observables, which contain large
long-distance contributions. For kaons, the eigenstates are usually
parameterized in terms of $\bar\varepsilon=(1+q/p)/(1-q/p)$ and different
approximations hold, see for example ref.~\cite{hep-ph/9911321}
for details on kaon mixing.

Finally, the third manifestation of CP violation in meson decays is through
the interference between mixing and decay. The key observable in this case
is the time-dependent CP asymmetry of a meson state $M(t)$ decaying into a final
state $f$. Taking a CP-eigenstate final state and up to $\mathcal{O}(\vert q/p\vert^2
-1)$ corrections, the asymmetry is given by
\begin{eqnarray}
  \label{eq:acp}
    A_\mathrm{CP}^{M \to f}(t)&=&\frac{\Gamma(M(t)\to f)-\Gamma({\bar M}(t)\to
    f)}{\Gamma(M(t)\to f)+\Gamma({\bar M}(t)\to f)}\nonumber\\
    &\simeq& \frac{C_{M\to f}\cos(\Delta m\, t) - S_{M\to f} \sin(\Delta m\, t)}
       {\cosh(\frac{\Delta\Gamma}{2}t)+S^\prime_{M\to f}\sinh(\frac{\Delta\Gamma}{2}t)}\,,
\end{eqnarray}
where 
\begin{equation}
  \label{eq:SandC}
  C_{M\to f}=
    \frac{1-\vert\lambda_{M\to f}\vert^2}{1+\vert\lambda_{M\to f}\vert^2}\,,~~
  S_{M\to f}=\frac{2\,\mathrm{Im}(\lambda_{M\to f})}{1+
    \vert\lambda_{M\to f}\vert^2}\,,~~ 
  S^\prime_{M\to f}=\frac{2\,\mathrm{Re}(\lambda_{M\to f})}{1+
    \vert\lambda_{M\to f}\vert^2}\,,
\end{equation}
with
\begin{equation}
  \label{eq:lambda}
  \lambda_{M\to f}=(q/p)_M\frac{\bar A}{A}\,.
\end{equation}
$(q/p)_M$ is the $M$ mixing parameter and $A$ ($\bar A$)
is the amplitude for $M\to f$ ($\bar M\to f$).
The coefficient $C_{M\to f}\simeq -A_\mathrm{CP}$, the direct CP asymmetry defined
in eq.~(\ref{eq:acpdir}). The coefficient $S_{M\to f}$, instead, is a new measurement
of $CP$ violation generated by the interference of mixing and decay amplitudes.
% Notice, in fact,
% that neither $q/p$ (mixing) nor $\mathcal{\bar A}/\mathcal{A}$ (decay) are
% separately observable, but the product $\lambda$ in eq.~(\ref{eq:lambda}) is.

Time-dependent CP asymmetries for $B$ decays (where $\Delta\Gamma\sim 0$) give access
to the UT angles. For example, as the amplitude for $B_d\to J/\psi K_S$ is dominated by
a single term with a definite weak phase, one finds
% for the measurement of the UT angle $\beta$. The final state
% is a CP eigenstate with eigenvalue $-1$, $\Delta\Gamma\sim 0$ for $B_d$ and the decay
% amplitude is dominated by a single term with a definite weak phase. One then finds
% \begin{equation}
% \lambda_{B_d\to J/\psi K_{S,L}}
% \simeq -\left(\frac{V_{tb}^*V_{td}}{V_{cb}^*V_{cd}}\right)
% \left(\frac{V_{cb}V_{cd}^*}{V_{tb}V_{td}^*}\right)=-e^{-2i\beta}\,,
% \label{eq:lambdaSL}
% \end{equation}
% so that
\begin{equation}
\label{eq:SC0}
A_\mathrm{CP}^{B_d\to J/\psi K_S}(t)=-S_{B_d\to J/\psi K_S}\sin(\Delta m\, t)\,,\quad
S_{B_d\to J/\psi K_S}\simeq\sin 2\beta\,,
\end{equation}
up to doubly Cabibbo-suppressed corrections.
In other cases, the decay amplitude is not dominated by a single term and
the extraction of the UT angles is less straightforward, as hadronic amplitudes
no longer cancel in $\lambda_{M\to f}$.
Previous formulae can also be generalized to non CP-eigenstate final
states~\cite{Dunietz:1986vi}.

\section{Present Status of Flavor Physics}
\label{sec:status}
In this section, we briefly review the present status of flavor physics in the SM,
discussing the determination of the CKM parameters and possible deviations
found in present data.

\subsection{Determination Of The Cabibbo-Kobayashi-Maskawa Matrix}
\begin{figure}[h!]
  \begin{center}
    \centerline{\includegraphics[width=0.53\textwidth]{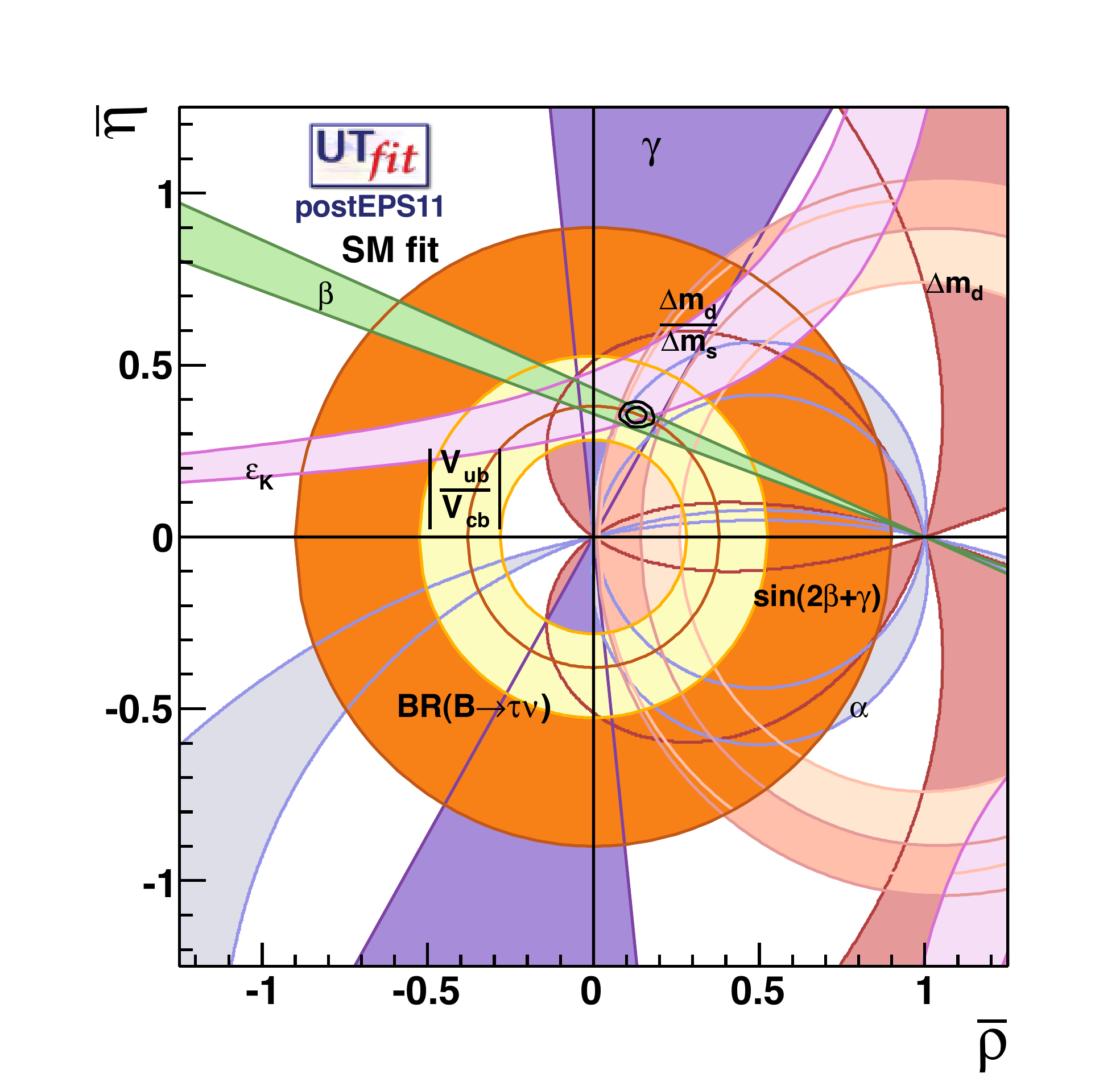}}
    \caption{Determination of $\bar\rho$ and $\bar\eta$ selected from constraints on
     $|V_{ub}/V_{cb}|$, $\Delta m_d$, $\Delta m_s$, $\varepsilon_K$, $\alpha$, $\beta$, $\gamma$,
     $2\beta$+$\gamma$ and  BR$(B \to \tau \nu)$.
     $68\%$ and $95\%$ total probability contours are shown, 
     together with $95\%$ probability regions from the individual constraints.
   }
    \label{fig:ckm_fit_today}
  \end{center}
\end{figure}

The CKM matrix elements $|V_{ud}|$ and $|V_{us}|$ can be measured
from super-allowed $\beta$ decays~\cite{Towner:2010zz} and semileptonic/leptonic
kaon decays~\cite{Antonelli:2010yf} respectively, determining accurately the sine of the
Cabibbo angle. The other CKM parameters are determined through a fit to the UT in
eq.~(\ref{eq:UT}), as shown in fig.~\ref{fig:ckm_fit_today}, using the latest determinations
of the theoretical and experimental parameters~\cite{Bevan:2010gi}.
The basic constraints are $|V_{ub}/V_{cb}|$ from semileptonic
$B$ decays, $\Delta m_d$ and $\Delta m_s$ from $B_{d,s}$ oscillations,
$\varepsilon_K$ from $K^0$--$\bar K^0$ mixing, $\alpha$ from charmless hadronic
$B$ decays, $\gamma$ and $2\beta$+$\gamma$ from charm hadronic $B$ decays,
$\sin2\beta$ from $B^0\to J/\psi K^0$ decays and the BR$(B \to \tau \nu)$~\cite{Asner:2010qj}.

%\epsfbox[0 0 30 50]{fig1.ps}
% %\epsfscale1200         % Figure enlarged to 120 (MAC)%
% \epsfxsize10pc         %

The consistency between the different constraints clearly establishes 
the CKM matrix as the dominant source of flavor mixing and CP-violation, 
described by a single parameter $\bar\eta$.

\subsection{Possible Hints of New Physics}

% \begin{figure*}[!tb]      
%     \includegraphics[width=0.48\textwidth]{pull_sm_sin2bfit3.eps}
%     \includegraphics[width=0.48\textwidth
% ]{figures/pull_sm_btaunufit8.eps}
%   \caption{The compatibility between the direct and indirect 
%     determinations of $\sin 2\beta$ (left) BR($B \to \tau\nu$) (right)
%     as a function of the measured values and errors.
%     The compatibility regions from $1\sigma$ to $6\sigma$
%     are displayed. The crosses display the position (value/error) 
%     of the measurements. }
%   \label{fig:pulls}
% \end{figure*}
% The consistency of the CKM picture can be quantitatively tested using 
% compatibility plots comparing two different probability density functions (pdf's):
% the one obtained from 
% the UT fit without using the constraint being tested and the other from the 
% direct measurement.
% The number of standard deviations between the measured value and the predicted
% value is plotted as a function of the measured value and its error. The
% compatibility can be then directly estimated on the plot, for any central
% value and error of the direct measurement. 

The consistency of the CKM picture can be quantitatively tested comparing the
direct measurement of a flavor observable entering the UT fit with the SM prediction
obtained from the UT fit without using the constraint being tested.

Table~\ref{tab:pull} contains a set of flavor observables: for each of them,
the SM prediction, the measurement, and the
pull (the difference of prediction and measurement in unit of $\sigma$) are shown.
At present, the main tensions in the UT fit come from BR$(B\to\tau\nu)$,
which is found to deviate from the measurement by $\sim 3.2\sigma$,
and $\sin 2\beta$, which is larger than the experimental value by
$\sim 2.6\sigma$. 
It is interesting to note that the former prefers 
large values of $|V_{ub}/V_{cb}|$, while the latter wants $|V_{ub}/V_{cb}|$ 
to be small. Any change of the measured value of $|V_{ub}|$ and $|V_{cb}|$ 
will not improve the situation.

In the $B_s$ sector, the possibility of a large deviation in $\beta_s$
is still open, although new Tevatron data on $B_s\to J/\psi\phi$, still to
be included in the average~\cite{tevav}, go in the SM direction~\cite{newCDF,newD0}. 
Moreover, the dimuon asymmetry $A_{\mu\mu}$ measured by D\O~\cite{Abazov:2010hv}
also displays a large deviation
from the SM, pointing to a large $\beta_s$ and possibly to NP contributions
in $\Delta\Gamma_s$.

\begin{table}[h!]
\caption{Summary of different measurements and corresponding SM 
predictions. In the last column the pull is explicitly indicated.\label{tab:pull}}
\centering
{\small
\begin{tabular}{|l|c|c|c|}
\hline
Observable & Prediction & Measurement & Pull ($\sigma$)  \\
\hline
$\gamma\, [^\circ]$ & $69.6\pm 3.1$ & $74\pm 11$ & $-0.4$ \\
$\alpha\, [^\circ]$ & $85.4\pm 3.7$ & $91.4\pm 6.1$ & $-0.8$ \\
$\sin2\beta$ & $0.771\pm 0.036$ & $0.654 \pm 0.026$ & $+2.6$ \\
$|V_{ub}|\,[10^{-3}]$ & $3.55 \pm 0.14$ & $3.76 \pm 0.20$ & $-0.9$ \\
$|V_{cb}|\,[10^{-3}]$ & $42.69 \pm 0.99$ & $40.83 \pm 0.45$ & $+1.6$ \\
$\varepsilon_{K}\,[10^{-3}]$ & $1.92 \pm 0.18$ & $2.23 \pm 0.010$ & $-1.7$ \\
$BR(B\to\tau \nu)\,[10^{-4}]$ & $0.805\pm 0.071$ & $1.72\pm 0.28$ & $-3.2$ \\
$\Delta m_s\,[\mathrm{ps}^{-1}]$ & $17.77\pm 0.12$ & $18.3\pm 1.3$ & $-0.4$ \\
$\beta_s[^\circ]$ & $1.08\pm 0.04$ & Tevatron & \multirow{2}{2cm}{\centering$+2.1$} \\
$\Delta\Gamma_s\,[\mathrm{ps}^{-1}]$ & $0.11\pm 0.02$ & average &  \\
$A_{\mu\mu}\,[10^{-4}]$ & $-1.7\pm 0.5$ & $-95.7\pm 29.0$ & $+3.2$ \\ 
\hline
\end{tabular}
}
\end{table}

\section{Opportunities From Precision Flavor Physics}
\label{sec:opportunities}
\subsection{B Physics}
\subsection{CP Violation}
In this section we describe the golden modes related to CP violation
measurements. 
Among the CP-violating observable there are the UT triangle angles $\beta$, 
$\alpha$, $\gamma$ and $\beta_s$, the mixing-induced CP violation in
charmless hadronic $B$ decays, in $b \rightarrow s \gamma$ transitions
and  the direct CP violation in $b \rightarrow s \gamma$ and 
$b\rightarrow s \ell \ell$ decays.

\subsubsection{CP violation in $b\to c\bar cs$ transition: angles $\beta$ and $\beta_s$}
The mixing induced CP asymmetry in $B \rightarrow J/\psi K_S^0$ transition allows
to measure the angle $\beta$ of the UT in a theoretically clean way. The 
measurement of this CP asymmetry was the golden mode for the past generation of
$B$-factories which finally measured $\sin2\beta=0.654\pm 0.026$~\cite{Bevan:2010gi}.
In Section~\ref{sec:status} we have shown that this parameter deviates $2.6\sigma$ 
from the SM. Thus Iimproving the precision of this measurement is particularly important.  
Its present error is still dominated by a statistical component of
about $0.010$.  If this measurement is performed in a $B$-factory-like environment, 
the statistical precision will match the systematic one for an integrated luminosity of
about 10 ab$^{-1}$.
Some preliminary study has shown that a better understanding of the detector
and the use  of control sample on data could allow to reduce the systematics to get a final 
experimental error of about $0.005$.
At this level of precision the theoretical error matters yet it is possible to control it
via a data-driven approach by measuring the time-dependent CP asymmetry in
$B^0\to J/\psi\pi^0$~\cite{Ciuchini:2005mg}.
%\cite{Ciuchini:2005mg,Faller:2008zc}.
One can conclude that it is possible to 
discover NP (at 5$\sigma$) if there is a deviation of $0.02$ from the SM expectation
for $\sin2\beta$ as measured in tree decays~\cite{O'Leary:2010af}.
Notice that the precision of this measurement is not expected to improve much at
hadronic machines.

Similarly, the mixing induced CP asymmetry in $B_s\to J/\psi \phi$ 
allows to measure the angle $\beta_s$. The SM value for this angle is 
very small, about $0.02$. This analysis has already been performed at the Tevatron and
will be one of the golden channels of the LHCb physics program.
%It cannot be efficiently performed 
%with $B_s$ produced at the $\Upsilon(5S)$ resonnance beacuse this analysis 
%requires to reconstruct the fast $B_s$ oscillation. An additional difficulty 
%is that the final states involve two vector mesons and so a full anglular 
%analysis is needed.
CDF and D\O\ have performed this analysis finding $\beta_s$ in $[0.02,0.52]
\bigcup [1.08,1.55]$ at $68\%$ C.L.~\cite{newCDF} and
$-2\beta_s=-0.76^{+0.38}_{-0.36}\pm 0.02$~\cite{newD0} respectively.
The Tevatron average, available only for older data, gives
$\beta_s$ in $[0.27, 0.59] \bigcup [0.97, 1.30]$ at $68\%$ C.L.~\cite{tevav}

LHCb should reach a sub-degree ($0.5^\circ$) precision with about 10 fb$^{-1}$. 
At this level of precision the theoretical error matters, but it can be controlled with
data-driven methods~\cite{Faller:2008gt}. Careful studies of systematics errors have to
be performed to prove that the precision can be further improved
with a larger data sample (LHCb upgrade).

\subsubsection{CP violation in $b\to s\bar sq$ transition}
The measurement of the mixing-induced CP violation in the $b\to s\bar sq$ 
is interesting, since new particles in the loop can cause deviations from 
SM predictions. In the SM these decays measure $\sin2\beta$ up to
channel-dependent corrections. Many channels have been measured at the 
$B$-factories and the ``golden four'' theoretically cleanest
ones are~\cite{O'Leary:2010af}:
$B \to (\phi K^0)$ $B \to \eta^\prime K^0$, $B^0 \to f_0 K^0$, 
$B \to K^0 K^0 K^0$. For these channels, $\Delta S = S(css)- S( J/\psi K_S^0)$ 
has been evaluated using hadronic models based on
%factorization~\cite{Beneke:2005pu,Cheng:2005bg,Cheng:2005ug}
factorization~\cite{Beneke:2005pu}
and found to be small ($\sim 0.02$) with an uncertainty of about $0.02$. 
However we stress that not all sources of theoretical error are under 
control in these estimates. On the other hand, data-driven methods can benefit from 
precision measurements so that other modes, such as $B\to K_S \pi^0$~\cite{Ciuchini:2008eh},
could become competitive at next-generation experiments.

From the experimental point of view, it has been shown that 
the precision of $S$ measured for the ``golden four'' can be pushed down to about 
$0.02$ ($0.01$ in case of $\eta^\prime K^0$) with about 75 ab$^{-1}$.
In case a sizable $\Delta S$, between 
$0.05$ and $0.10$, is found in these measurements, further theoretical and phenomenological 
work will be required to pin down the SM value and firmly establish the presence of NP. 
In this respect, the opportunity of measuring several modes with different theoretical 
uncertainties, but possibly correlated NP contributions, is a unique advantage of
super flavor factories.

In this panorama, LHCb can contribute with one measurement, $S(B_s\to \phi\phi)$,
reaching a precision of about $0.03$.

\subsubsection{CP violation in charmless $B \rightarrow \pi(\rho) \pi(\rho)$ decays: angle $\alpha$}
The angle $\alpha$ can be obtained using the time-dependent analyses of 
$B^0 \to \pi^+ \pi^-$, $B^0 \to \rho^+ \rho^-$ and $B^0 \to (\rho \pi)^0$.                
These charmless decays are dominated by the $b \to u$ tree amplitude ($T$).
Neglecting the second amplitude with a different weak phase, the
so-called penguin amplitude ($P$), these decays give a measurement of $\sin 2\alpha$.
Actually the experimentally measured quantity is $\sin 2\alpha_{e\!f\!f}$, 
which is a function of $\sin 2\alpha$ but also of the unknown hadronic parameter
$P/T$. Several strategies have been proposed to get rid of this
``penguin pollution''. The original method uses an analysis of all the
$B\to\pi\pi$ decays based on the $SU(2)$ flavor symmetry~\cite{Gronau:1990ka}.
The $B$-factories have measured $\alpha= (91.4 \pm 6.1)^\circ$.
With the full dataset, super flavor factories will measure $\alpha$ with precision of 
$(2-3)^{\circ}$ separately for $B^0 \to \pi^+ \pi^-$, 
$B^0 \to \rho^+ \rho^-$ and $B^0 \to (\rho \pi)^0$ decay modes. At this level of accuracy,
isospin violation may become relevant. However, super flavor factories permit
a multi-approach strategy allowing to test the consistency among results for $\alpha$
obtained in different channels.

The measurement of $\alpha$ will be extremely difficult in a hadronic environment
(but for one measurement at about 5$^{\circ}$ in the $\rho\pi$ channel) since it
typically requires the reconstruction of decay modes containing several neutral particles.

\subsubsection{CP violation in $B \rightarrow D K$ transition: angle $\gamma$}
Various methods related to $B \to DK$ decays have been proposed to determine 
the UT angle
%$\gamma$~\cite{Gronau:1990ra,Gronau:1991dp,Dunietz:1991yd,Dunietz:1992ti,Atwood:1994zm,Atwood:1996ci,Giri:2003ty},
$\gamma$~\cite{Gronau:1990ra,Dunietz:1991yd,Atwood:1994zm,Giri:2003ty},
using the fact that 
a charged $B$ can decay into a $D^0(\overline{D}^0) K$ final state via a 
$V_{cb}$($V_{ub})$ mediated process.  CP violation occurs if $D^0$ and 
$\overline{D}^0$ decay to the same final state.  
These processes are thus sensitive to the phase difference $\gamma$ between 
$V_{ub}$  and $V_{cb}$. The same argument can be applied to $B \to D^{*}K$ and 
$B \to D^{(*)} K^*$ decays and for neutral $B$ decays such as $B^0 \to 
D^0(\overline{D}^0) K^{*0}$.

The angle $\gamma$ has been measured at $B$-factories with an 
unexpected precision : $\gamma = (74 \pm 11)^\circ$ with a $180^\circ$ ambiguity. 
Using all the different methods and the large data set available from 
next-generation experiments, the precision on $\gamma$ can be improved down to
the degree level, where the precision will be still dominated by the statistical error. 
It has to be noted that LHCb is expected to reach a precision on $\gamma$ of 
about $(2-3)^\circ$ with about 10 fb$^{-1}$.

\subsubsection{CP violation in radiative $b\to s\gamma$ decays}
The study of the direct CP asymmetry in radiative $B$ decays is one of the 
golden modes of the physics programme at super flavor factories.
This asymmetry is predicted very small, of order $0.5\%$, in SM
and with a relatively small theoretical uncertainty.
The exclusive decays $A_{CP}(B \to K^* \gamma)$ have been actually 
measured at the present $B$-factories and is compatible with zero
within an error  of about $0.025$. The present error is already limited
by systematic effects due to asymmetries in the detector response
to positive and  negative kaons. Recent studies has shown the use of
large data control sample could allow to push the precision down to $0.5\%$.
The inclusive decays have been also studied at the present $B$-factories
and $A_{CP}$ found compatible with zero. Also in this case the error with the $75$
ab$^{-1}$ could be pushed down to $0.5\%$.

In the SM,
%the partial widths differences in $b \to s \gamma$ 
%and $b \to d \gamma$ cancels and thus 
the asymmetry $A_{CP}(B\to X_{d+s} \gamma)$ is expected to vanish.
From the experimental point of view, the study
of this sample is interesting since the systematics from the particle
identification does not contribute. On the other hand one has to 
deal with a larger background. Preliminary studies have shown that 
a sub-percent precision is reachable with 75 ab$^{-1}$ at the super
flavor factories.

\subsubsection{CP violation in radiative $b\to s\ell^+\ell^-$ decays}
The exclusive mode $B \to K^* \mu^+\mu^-$ will be deeply studied by LHCb
and several observable precisely measured. Nevertheless, these exclusive 
measurements will be limited by hadronic uncertainties. 
The role of next-generation experiments in this sector is to provide
theoretically cleaner measurements of the same observables
using inclusive modes.
Recent studies has shown that a sample exceeding $10$ ab$^{-1}$ is needed
to take advantage of the theoretical cleanliness of several inclusive 
observables, such as the zero-crossing of the forward-backward asymmetry 
in $b \to s\ell^+\ell^-$~\cite{O'Leary:2010af}.

\subsubsection{CP violation in radiative $b\to s \gamma$ decays 
and photon polarization}
Within the SM, the photon emitted in radiative $b$ ($\bar b$) decays are 
predominately left (right)-handed. The measurement of the mixing-induced 
CP asymmetry allows to probe the interference between $b$ and 
$\bar b$ decays and thus the polarization of the photon.
In the SM the value of the parameter $S$ is expected to be below 5$\%$.
The current $B$-factories have measured $S(B\to K_S \pi^0 \gamma)$
with about 25$\%$ precision.
This measurement can be converted to an uncertainty of about $0.16$ on the 
fraction of the wrongly-polarized photons ($R$(wrong)). The uncertainty is dominated 
by the statistical error. The critical point for this measurement is the 
efficiency in the reconstruction of the $K_S^0$ which depends crucially 
on the radius of the silicon detector.
Recent sensitivity studies have shown that the  error on $S$ can be pushed down to $0.02$
and thus to about $0.02$--$0.03$ on $R$(wrong) with a dataset of 75 ab$^{-1}$.

Similar analyses can be performed by LHCb using $B\to \phi \gamma$.
Using the full dataset, a precision of about $0.1$ on $R$(wrong) can be achieved.
It is worth noting that $B \to K^* e^+ e^-$ can give a competitive measurement
of $R$(wrong).

\subsection{Rare Decays}
\subsubsection{$B \to K^{(*)} \nu \bar \nu$ decay modes}
The rare decay $B \to K^{(*)} \nu \bar \nu$ is an
interesting probe of NP in $Z^0$ penguins~\cite{Buchalla:2000sk}, 
as for example chargino-up-squark contributions in a generic supersymmetric theory. 
The $b\to s\nu\bar\nu$ transition is governed by the effective Hamiltonian
\begin{equation}
{\mathcal H}_{\rm eff} = - \frac{4\,G_F}{\sqrt{2}}V_{tb}V_{ts}^*\left(C^\nu_L \mathcal 
O^\nu_L +C^\nu_R \mathcal O^\nu_R  \right) + {\rm H.c.}\,,
\end{equation}
where the operators are $\mathcal{O}^\nu_{L,R} =\frac{e^2}{8\pi^2}(\bar{s}  
\gamma_{\mu} P_{L,R} b)(  \bar{\nu}\gamma^{\mu}  P_L \nu)$,
and the $C^\nu_{L,R}$ are the corresponding Wilson coefficients. In the 
SM, $C^\nu_L \approx - 6.38$ and the $C^\nu_R$ vanishes.
%In models beyond the SM, both $C^\nu_L$ and $C^\nu_R$ can be non-zero and complex.
% However, the two exclusive and the inclusive 
% decay rates, as well as the longitudinal polarization fraction $F_L$ in the channel
% with $K^*$,
Observables in $B \to K^{(*)} \nu \bar \nu$ decays
only depend on two independent combinations 
of these Wilson coefficients, which can be written as~\cite{Altmannshofer:2009ma}
\begin{equation}  \label{eq:epsetadef}
 \epsilon = \frac{\sqrt{ |C^\nu_L|^2 + |C^\nu_R|^2}}{|(C^\nu_L)^{\rm SM}|}~, \qquad
 \eta = \frac{-{\rm Re}\left(C^\nu_L C_R^{\nu *}\right)}{|C^\nu_L|^2 + |C^\nu_R|^2}~,
\end{equation}
with $(\epsilon,\eta)_{\rm SM}=(1,0)$.

Due to presence of two undetected neutrinos the analyses of these decay modes are 
particularly challenging. Recent studies have shown that the $3\sigma$ observation of 
BR($B \to K \nu \bar \nu$) is expected with a data sample of 10 ab$^{-1}$
at super flavor factories, while 50 ab$^{-1}$ will be needed to observe
$B \to K^* \nu \bar \nu$, assuming the SM decay rates.
In addition the longitudinal polarization fraction $F_L$ for $B \to K^* \nu \bar \nu$ can
also be measured  with the decent precision of about 20$\%$. 

\begin{figure}[h!]
  \begin{center}
    \includegraphics[width=0.6\textwidth]{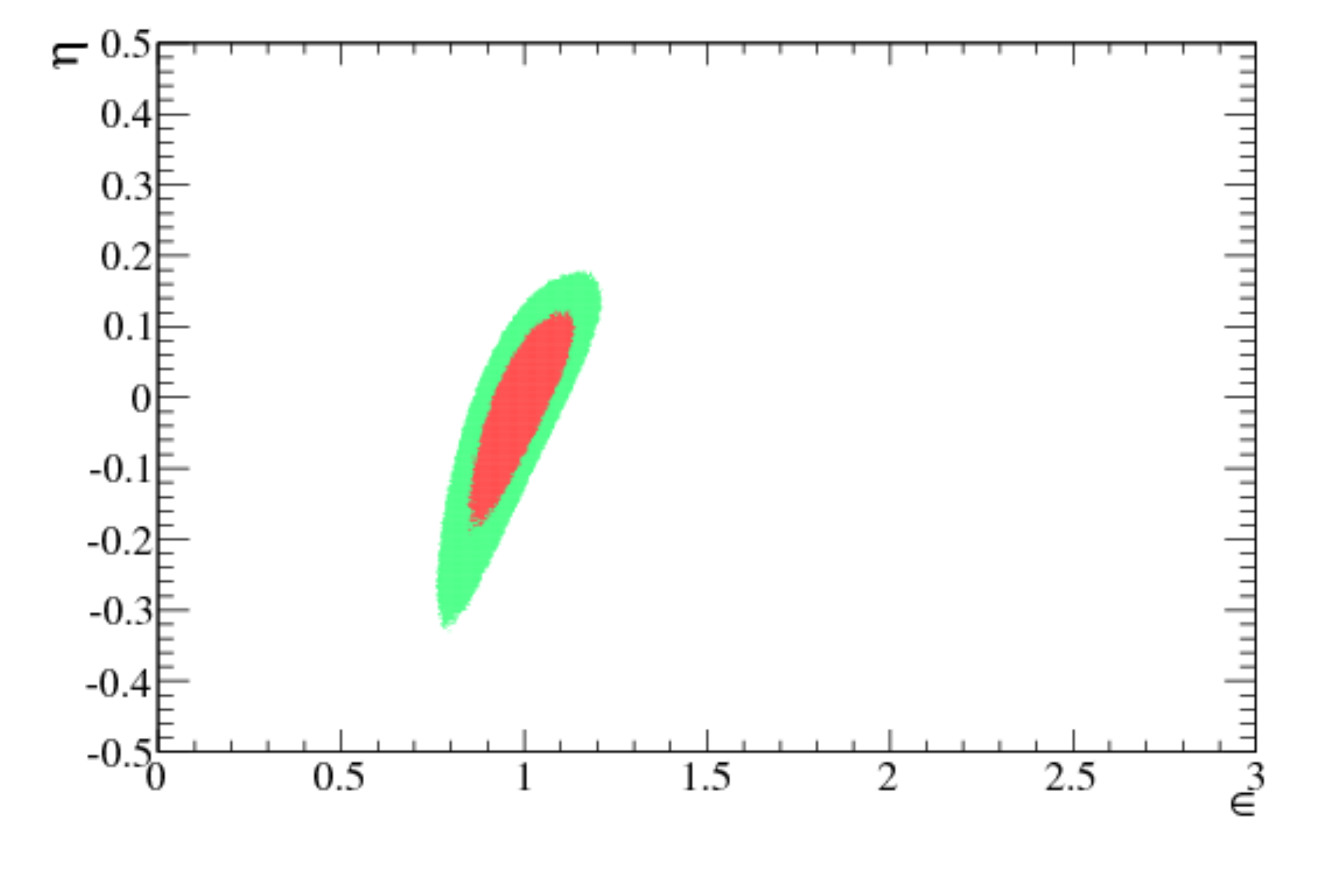}
  \caption{\label{fig:angular_constraint} Expected constraint on the $(\epsilon,\eta)$ 
plane from the measurement of BR($B \to K^{(*)} \nu \bar \nu$) 
and the angular analysis of $B^0 \to K^{*0} \nu \bar \nu$ with a dataset of 75 ab$^{-1}$. }
  \end{center}
\end{figure}

The combination of this information with the measurement of the branching ratios 
would provide a constraint in the plane $(\epsilon,\eta)$, as shown in
Fig.~\ref{fig:angular_constraint}, where NP would show up as a deviation from the
SM values $(1,0)$.

\subsubsection{The leptonic branching fractions $B \to \ell \nu$}
\begin{figure}[!tb]
  \begin{center}
    \includegraphics[width=4.cm,angle=-90]{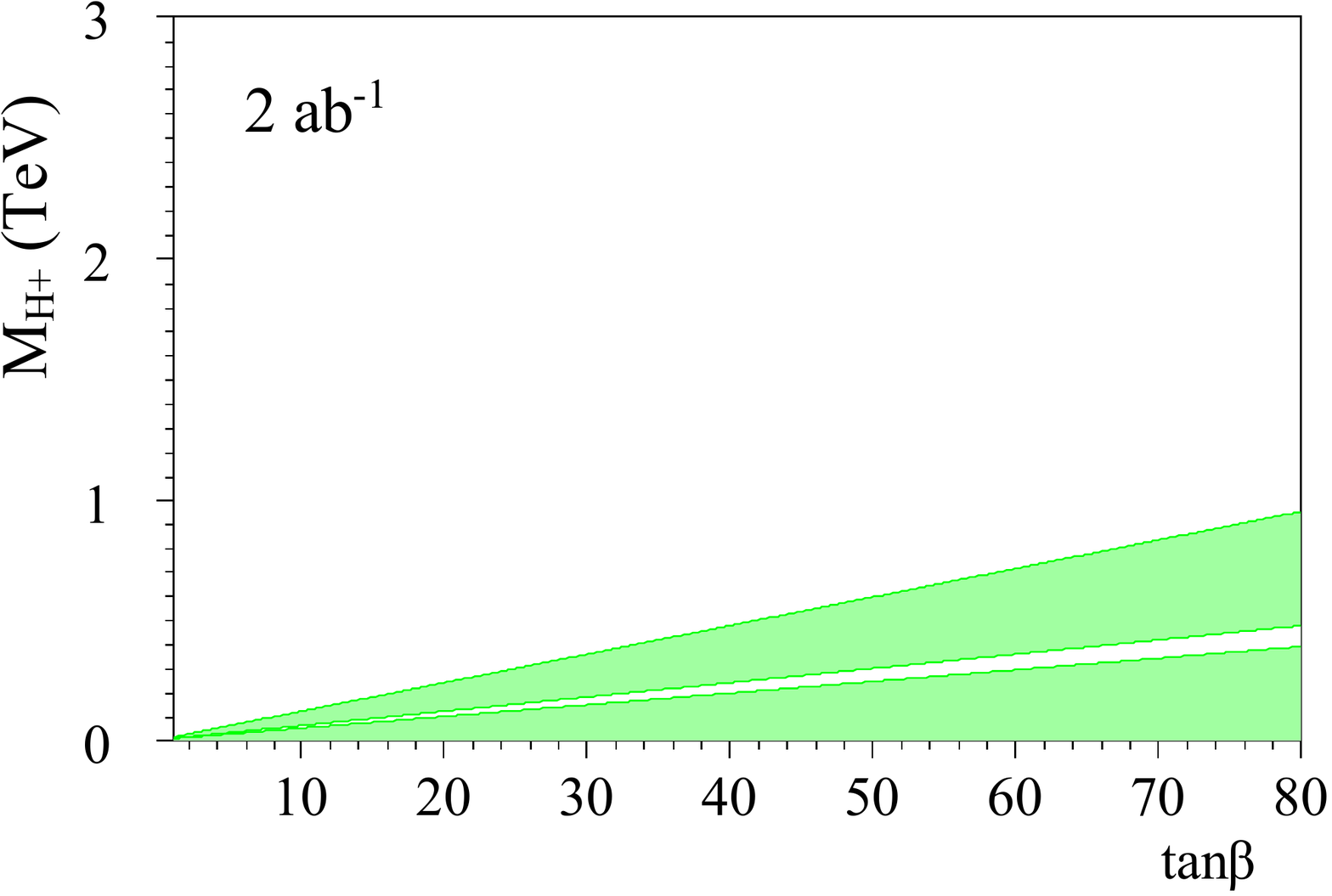}
    \includegraphics[width=4.cm,angle=-90]{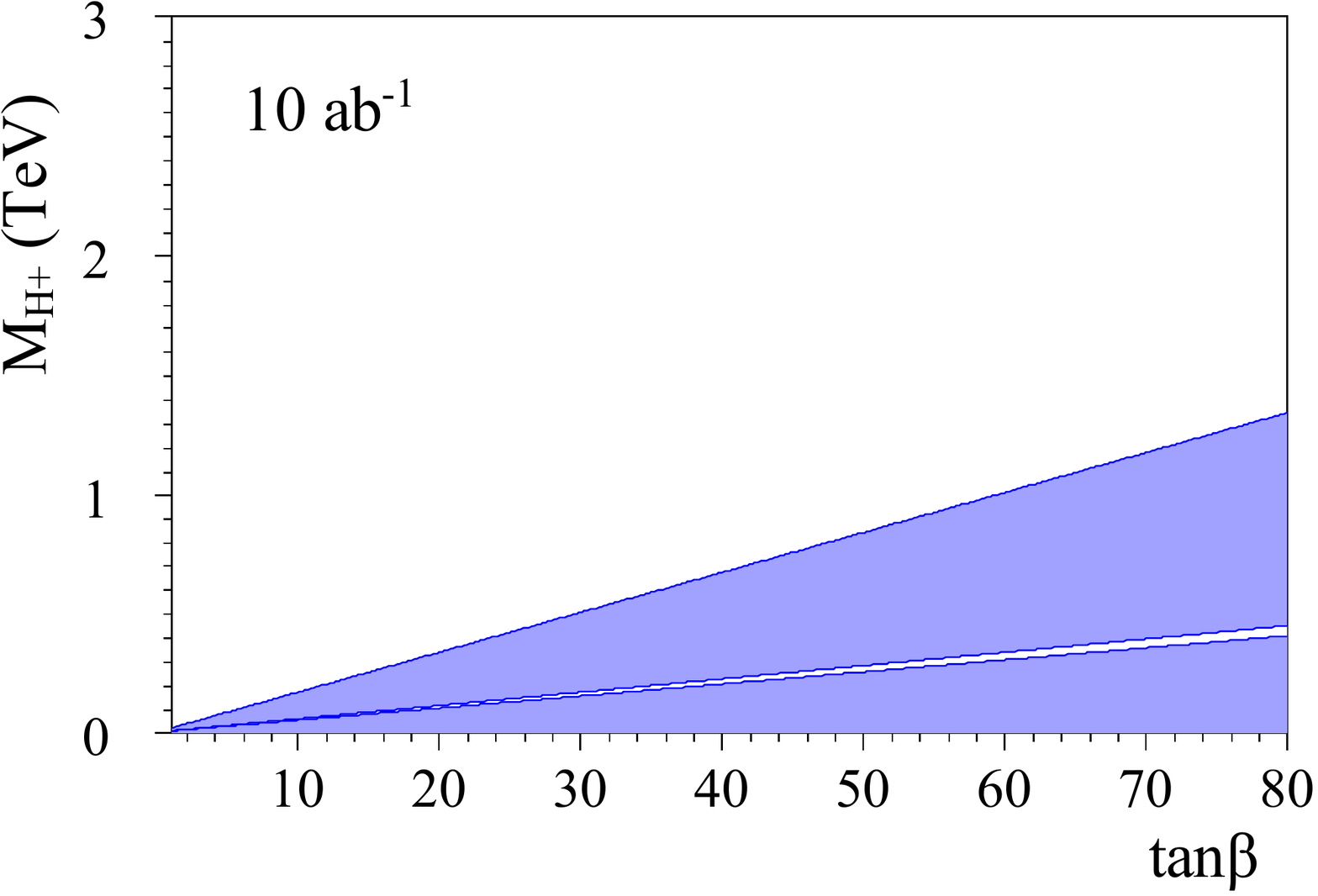}
    \includegraphics[width=4.cm,angle=-90]{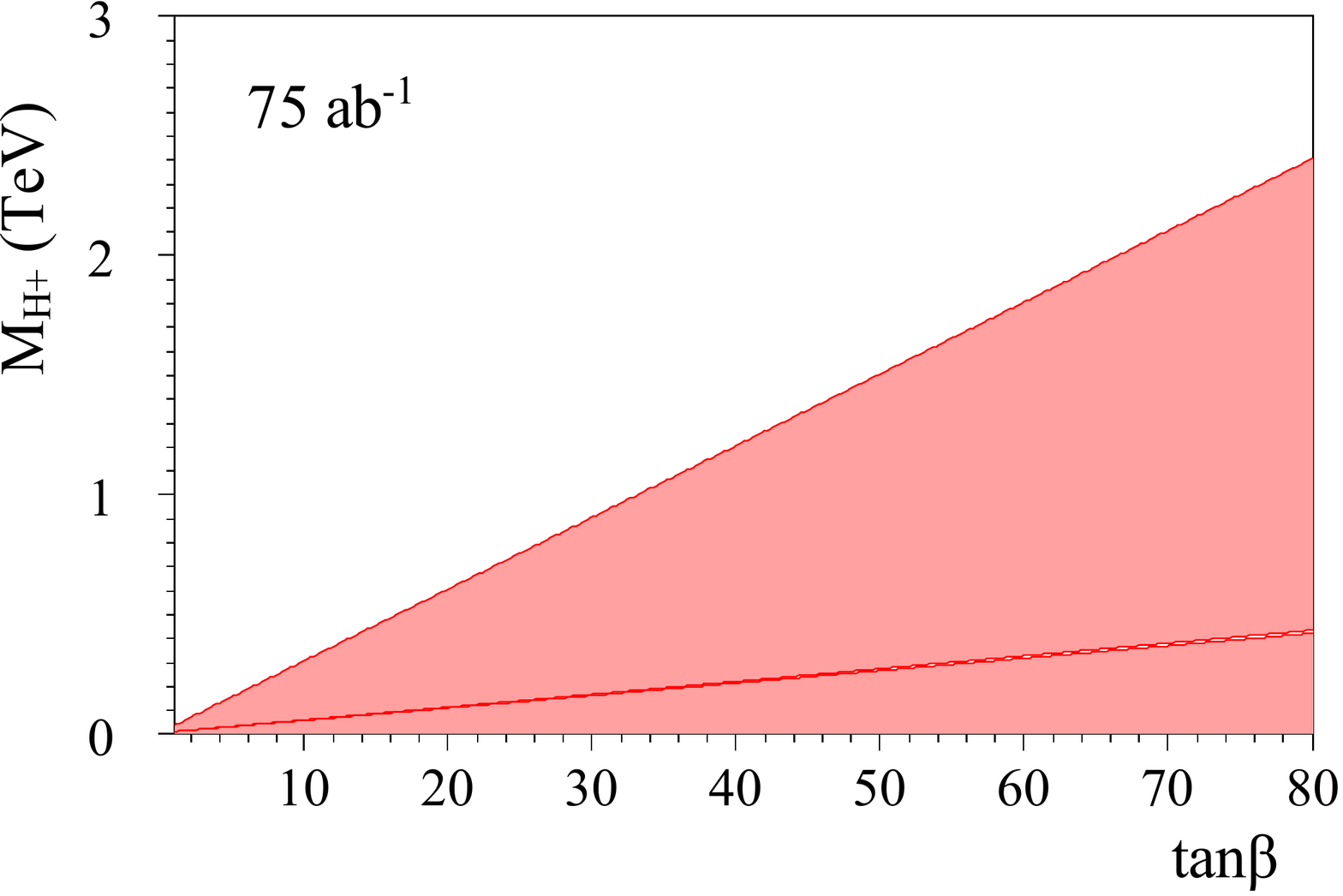}
    \includegraphics[width=4.cm,angle=-90]{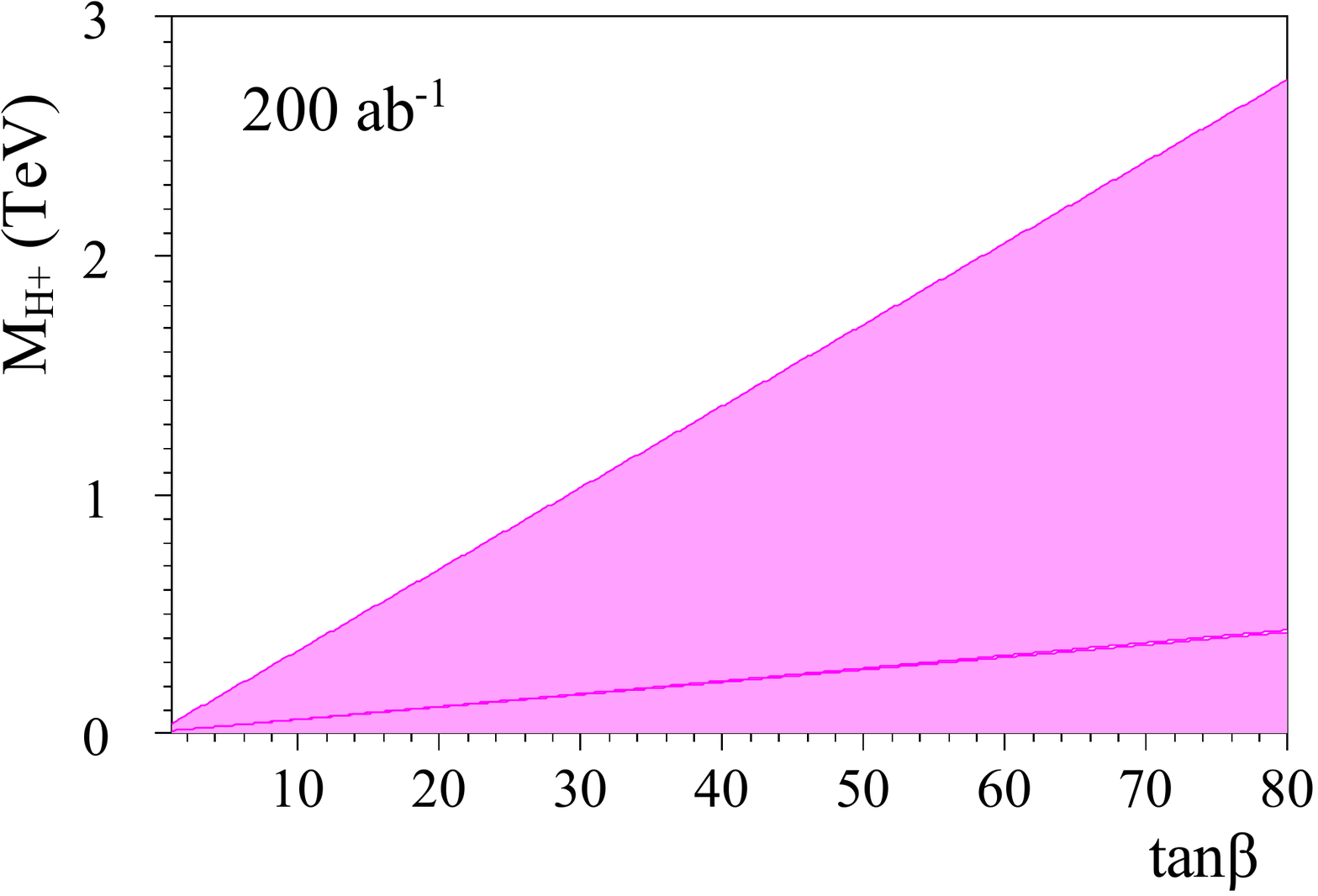}
    \caption{\label{fig:btaunu}
      Exclusion regions in the $m(H^+)$--$\tan\beta$ plane arising from the
      combinations of the measurement of ${\cal B}(B \to \tau \nu)$ and
      ${\cal B}(B \to \mu \nu)$ using 2 ab$^{-1}$ (top left), 10 ab$^{-1}$ (top right)
      75 ab$^{-1}$ (bottom left) and 200 ab$^{-1}$ (bottom right). We assume that
      the result is consistent with the Standard Model.
    }
  \end{center}
\end{figure}

Precision measurements of the branching fraction of $B\to \ell \nu$
where $\ell = e, \mu, \tau$  can be used to constrain NP contributions which could
enhance it, see for example ref.~\cite{Bona:2007qt}.
Fig.~\ref{fig:btaunu} shows a comparison of exclusion plots in the
$m(H^+)$--$\tan\beta$ plane of a Two-Higgs-Doublet Model (Type II) coming from a
measurement of BR$(B \to \tau \nu)$ and BR$(B \to \mu \nu)$ with different data samples,
2 ab$^{-1}$, 10 ab$^{-1}$, 75 ab$^{-1}$ and 200 ab$^{-1}$, 
assuming SM values for the BRs.

Moving from 10 ab$^{-1}$ to 75 ab$^{-1}$, the channel $B \to \mu \nu$,
which is experimentally cleaner, begins to give a significant contribution to the 
average, and the gain is then larger than the naive statistical expectation. Increasing
the integrated luminosity beyond 75 ab$^{-1}$,  $B \to \mu \nu$ 
overtakes $B \to \tau \nu$ which becomes systematic limited.

It is clear from fig.~\ref{fig:btaunu} that the presence of charged Higgs with mass
beyond the TeV could be detected in scenario with high $\tan\beta$.

\subsubsection{The leptonic $B \to \ell^+ \ell^-$}
The decay $B_s \to \mu^+ \mu^-$ suffers from loop and helicity suppressions
in the SM and is predicted to be extremely small, $BR(B_s \to \mu^+ \mu^-)$ =
$(3.7\pm 0.5)\times 10^{-9}$~\cite{Bona:2009cj}. NP extensions of the SM do not necessarily
share this suppression which can be enhanced by an order of magnitude or more.
The best current limit is $BR(B_s \to \mu^+ \mu^-)$ $<  4.3\times 10^{-8}$ at
95$\%$ C.L. obtained by CDF~\cite{ref:CDF_BsMumu}. 
The recent result from LHCb obtained using about $50$ pb$^{-1}$ gives already a very competitive limit:
$BR(B_s \to \mu^+ \mu^-)$ $< 5.6\times 10^{-8}$ at 95$\%$ C.L. \cite{ref:prelim}.
$B_s \to \mu^+ \mu^-$ is in fact a golden mode for LHCb which can overtake the sensitivity of CDF with 
about 100 pb$^{-1}$ and reach a precision of about $20\%$ with the full dataset of 
10 fb$^{-1}$. Further improvement requires the control of the $B_s$ production rate or the precise 
measurement of some reference $B_s$ absolute $BR$. The latter could be provided by a run 
of a super flavor factory at the $\Upsilon(5S)$ resonance. It is worth mentioning that, while
the LHCb measurement of BR($B_s \to \mu^+ \mu^-$) will seriously challenge the large $\tan\beta$
scenario, it cannot be excluded that large $\tan\beta$ effects show up in $B\to \ell \nu$
and not in $B_s \to \mu^+ \mu^-$.

\subsubsection{Semileptonic Decays}
The improvement of the measurement of the magnitude of the CKM matrix elements 
$|V_{cb}|$ and $|V_{ub}|$ can be only achieved at super flavor factories. 
In a landscape where $\gamma$ will be already measured with a good 
precision at LHCb, the determination of $|V_{cb}|$ and $|V_{ub}|$ is the key 
ingredient to precisely obtain the UT parameters $\bar\rho$ and $\bar\eta$ in the
presence of NP.
Two approaches are possible using inclusive and exclusive decays. 
The current precision for $|V_{ub}|$ using both approaches is about 10$\%$.
For the inclusive decays the use of a large sample of data allow to use the most 
clean analysis approaches (hadronic recoil tag) which have been already tested in 
the present $B$-factory. Recent studies have shown that an experimental error at 
about (2-3)$\%$ can be reached. How to improve the theoretical uncertainty on the
inclusive $|V_{ub}|$ determination at the percent level is an open issue.
If we assume that, in the super flavor factory era, this uncertainty will be dominated
by the knowledge of the $b$ quark mass, a total error at about 3$\%$ on $|V_{ub}|$
seems possible~\cite{Hitlin:2008gf}.
On the other hand, the measurement of $|V_{ub}|$ using exclusive decays is presently 
limited by theoretical uncertainties on the form factors (about 10$\%$). Lattice 
calculations are expected to significantly improve to approximately 2-3$\%$ in the 
case of the most promising decay $B \to \pi \ell \nu$~\cite{Bona:2007qt,O'Leary:2010af}.
From the experimental side, the use of the recoil tag technique will allow to keep the
total error at about 3$\%$. The two methods uses independent data sample and theory
uncertainties, thus an error on $|V_{ub}|$ below 2$\%$ could be possible.
Similarly $|V_{cb}|$ can be determined with both methods with a precision below the
percent level~\cite{O'Leary:2010af}.

\subsection{Charm Physics}
Charm physics could play an important role in the NP searches. In fact, among the up-type
quarks, only charm allows to probe FCNC (and thus NP) in oscillation phenomena and in particular 
those involving CP violation.
%Charm gives additional opportunities to look for large NP contributions
%to the up-type quark sector.
Note that, in the SM, direct CP violation in charm transitions 
only occurs in Cabibbo-suppressed modes at an observable level $\sim {\cal O}(10^{-3})$ and 
time dependent CP asymmetries could reach the $10^{-5}$ [$10^{-4}$] level in Cabibbo-allowed 
and singly[doubly]-suppressed modes. The recent observation of $D$--$\bar D$ oscillations,
with $x_D$, $y_D \simeq 0.005$--$0.01$, has clearly open the possibility of observing CP
violation which would be a clear manifestation of NP.

Super flavor factories  can perform studies on the charm sector in a comprehensive manner,
with a large data sample in the $\Upsilon(4S)$ region. SuperB can also run the $\psi(3770)$
resonance.  In fact the SuperB collider is designed to run at lower center-of-mass energies
and reduced luminosity. In addition, the option to have a boost sufficient to perform
time-dependent measurements is under study. With very short low-energy runs, a data sample
an order of magnitude larger than the final BES-III sample can be readily obtained. Running
at the charm threshold should allow to  measure precisely the $D$ decay form factors with
semileptonic decays and the $D$  decay constant with leptonic decays. These measurements
would provide important benchmarks for lattice QCD calculations. In addition, Dalitz
analyses with high statistics would provide inputs to the measurement of the UT angle
$\gamma$.  Finally, a run at the charm threshold could also be helpful for FCNC searches.

\subsection{$\tau$ Physics}
\subsubsection{Lepton Flavor Violation}
Lepton Flavor Violation (LFV) in $\tau$ decays is one of the most powerful
probe of NP. This search at super flavor factories is complementary
with the existing and future neutrino experiments aiming
at measuring $\theta_{13}$ and with the MEG experiment at PSI searching
for $\mu\to e \gamma$.
With an integrated luminosity of 75 ab$^{-1}$, SuperB can gain an order of
magnitude on several LFV in $\tau$ decays (see Table \ref{tab:lfv}),
exploring a significant portion of the parameter space of various
NP scenarios.
\begin{table}[!h]
\begin{center}
{\small
\begin{tabular}{c|c||c|c}\hline
 Final State     & Sensitivity $[10^{-10}]$ & Final State & Sensitivity $[10^{-10}]$\\ \hline \hline
 $\mu\gamma$     & 20 &  $\mu\eta$       & 4 \\
 $e\gamma$       & 20 &  $e\eta$         & 6 \\
 $3\mu$          & 2  & $\ell K^0_S$    & 2 \\
 $3e  $          & 2  & &
%  $\mu\eta$       & 4  \\
%  $e\eta$         & 6  \\
%  $\ell K^0_S$    & 2  \\
\end{tabular}
}
\end{center}
\caption{The experimental sensitivities (in units of $10^{-10}$) expected for LFV searches in $\tau$
decay with a dataset of 75 ab$^{-1}$.}\label{tab:lfv}
\end{table}

Super flavor factories are unique facilities for these studies. LHCb can contribute 
only to the 3$\mu$ channels with an estimated sensitivity smaller by about one order 
of magnitude. 

A longitudinally-polarized electron beam (at the level of about 85\%) can be obtained at SuperB and helps
to study the structure of LFV couplings in $\tau$ decays. Recent analyses have shown
that polarization gives new handles to discriminate between signal and background and thus probably 
allows to push even further the sensitivity of LFV measurements. 
Polarization is also important to search for $\tau$ EDM, $(g-2)_\tau$ and CP violation 
in $\tau$ decay.

\subsection{$K$ Physics}
The main issue of the Kaon physics in the next decade is the study of 
rare decays. The actors will be NA62 K$^+$-factory at CERN 
(and possibly P996 at Fermilab), KOTO $K_L$-factory at J-Park, KLOE-2 
$K_S$-factory at Frascati.

NA62 is approved and will start to take data at CERN in 2013 with the aim 
of collecting about 100 events $K^+ \to \pi^+ \nu \bar\nu$ using the in-flight 
technique. This would allow to reach a precision of $10\%$ on the 
determination of the $BR(K^+ \to\pi^+ \nu\bar \nu)$.
P996 at Fermilab could be constructed before 2020 and using the 
stopped-kaon technique will be able to reconstruct about 200 events.
KOTO experiment at J-PARC aims at reaching a single event sensitivity 
at the level of the SM expectation in three years running for the 
ultra rare decay $K_L \to \pi^0 \nu \bar\nu$.

For what concern the $K_S$ rare decays, KLOE-2 experiment, restarting
in 2011, aims at pushing down to 1$\%$ the error on $K_S \to \gamma \gamma$
and at observing few $K_S \to \pi^0 \ell \ell$ events with the
full dataset of 20 fb$^{-1}$.

It is worthwhile recalling that NA62 and KLOE-2 can render the test on lepton 
universality violation more stringent (few per mill) in the 
near future by studying the ratio between $K^+ \to e^+ \nu$ and $K^+ \to \mu^+ \nu$.
To reach the per-mill precision, however, next-generation experiments such as 
P36 at J-PARC are needed.
Finally KLOE-2 and NA62 can contribute in the searches for new light neutral 
boson associated with spontaneously broken global symmetries by studying
events like $K^+ \to \pi^+ X^0$.

\subsection{Other Opportunities}
\subsubsection{Spectroscopy and direct searches}
The recent results from the $B$-factories provide evidence for the
renaissance of hadronic spectroscopy. Although past performances
provide no guarantee of future success, new particles have been
discovered by the $B$-factories at a rate of
more than one per year, and there is no reason to believe that this
should not continue into the multi-ab$^{-1}$ territory.
Super flavor factories will open a unique window on this physics as
they allow for a high statistics study in a clean $e^+e^-$ environment,
ideal for the complicated analyses necessary to pin down the nature of these
new hadrons. Particles can be searched for in exclusive decays,
or by using inclusive techniques, such as the recoil analysis.
The possibility of running at different center-of-mass energies
(as at $\Upsilon(3S)$) extends the reach of this branch of the physics
programme.

The studies of lower $\Upsilon$ resonances would allow testing
extensions of the Standard Model in a manner complementary
to the physics program studied at previous $B$-factories and at the LHC.
Among the possibilities, we mention the search for a light pseudo-scalar Higgs
boson produced in the decay $\Upsilon(n{\rm S}) \to \ell\ell\gamma$ ($n=1,2,3$)
as an intermediate state occurring in models like the
%NMSSM~\cite{Fullana:2007uq,Domingo:2007dx,Ellwanger:2009dp}.
NMSSM~\cite{Fullana:2007uq}.
In addition, the study of the decays $\Upsilon(n{\rm S})$ to invisible 
allows to obtain independent constraints on models with light dark
matter~\cite{McElrath:2005bp}.

\subsubsection{Electroweak neutral current measurements}
With polarization, SuperB can measure the left-right asymmetry of $e^+e^-\rightarrow \mu^+\mu^-$
repeating at a lower energy the measurement performed by the SLC collaboration at the
$Z$-pole~\cite{Abe:1994wx,:2005ema}
SLC measured $\sin^2 \theta_{W}=0.23098\pm0.00026$. An important contribution to the error comes 
from systematic component of $\pm 0.00013$, dominated by the polarization uncertainty 
of $0.5\%$.
A feasibility study has been performed assuming a left-right asymmetry of $-0.0005$.
With 75 ab$^{-1}$ and $80\%$ polarization, the statistical error on the 
left-right asymmetry will be about $5\times 10^{-6}$ (a ${\cal O}(1\%)$ relative error). 
If the polarimeter systematic errors can be kept below this level, the uncertainty on  
$\sin^2 \theta_{W}$ will be $\sim 0.0002$, which is competitive with the SLC measurement.
Similar measurements can be made with $e^+e^-\rightarrow \tau^+\tau^-(\gamma)$ 
and with charm. In this case the larger errors are expected because of the lower selection 
efficiency.

These precision measurements are sensitive to the same new physics scenarios, such as a $Z^{\prime}$,
being probed by the QWeak experiment at the Jefferson Laboratory, which will measure
$\sin^2 \theta_{W}$ to approximately $0.3\%$ at  $Q^2=(0.16~\mathrm{GeV})^2$.
%Figure~\ref{fig:sin2thetarun} shows the current and planned measurements of $\sin^2 \theta_{W}$.
SuperB will provide a point at $Q^2=(10.58~\mathrm{GeV})^2$ with an error comparable to
that of the measurement at the $Z$-pole.
% \begin{figure}[tb!]
%   \begin{center}
%     \includegraphics[width=0.6\textwidth]{qweak_running.eps}
%     \caption{
% Summary of experiments that have measured or are proposing to measure $\sin^2 \theta_{W}$
% as compiled in \cite{Qweak}.
% The standard model  running of $\sin^2 \theta_{W}$ is overlaid on the data points.
% SuperB will provide a point
%  at $Q=10.58~\mathrm{GeV}$ with an error comparable to that of the measurement at the Z-pole. }
%         \label{fig:sin2thetarun}
%   \end{center}
% \end{figure}

\section{Beyond The Standard Model With Precision Flavor Physics}
\label{sec:bsm}
In this section we illustrate the impact of future flavor measurements on
a selection of NP models, profiting from the studies of
refs.~\cite{Bona:2007qt,Aushev:2010bq,O'Leary:2010af}.
Of course, the conclusions one can draw from this
exercise are strongly influenced by the status of particle physics
at the time these measurements are actually performed.
The framework for next-generation flavor physics is being
set by present experiments both in the flavor
sector and at high-$p_T$. NP signals in the near future
are certainly possible and eagerly awaited. Meanwhile,
for the sake of illustration, we consider various NP
models in different scenarios to show the role of precision
flavor physics in elucidating the structure of physics beyond
the SM. None of the considered models will be presented in detail.
We restrict our presentation to what is needed for our
purpose and refer the reader to the original literature for
any additional information.

\subsection{Precision CKM Matrix}
\begin{figure}[h!]
\centerline{\includegraphics[width=0.47\textwidth]{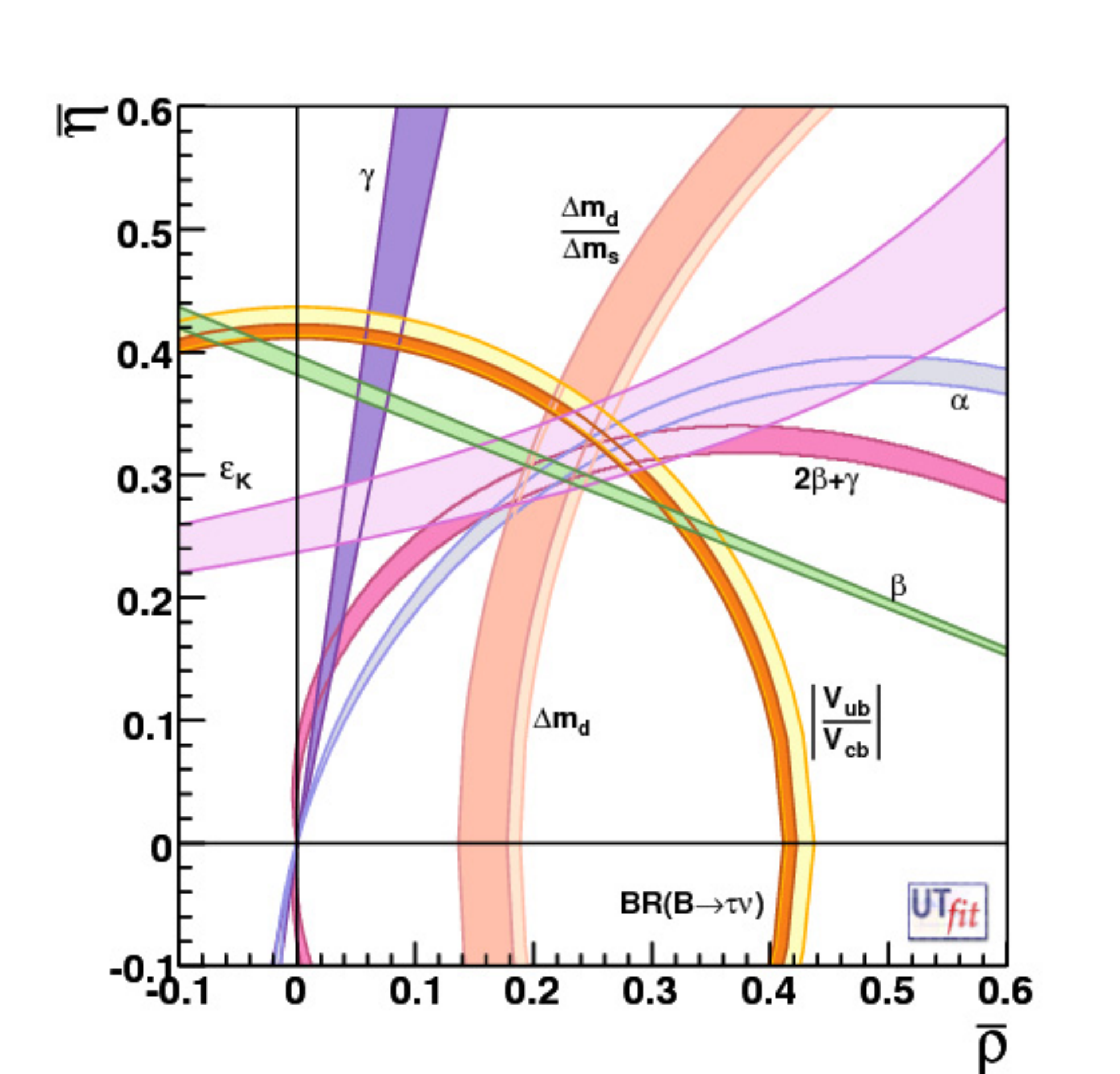}}
\caption{Extrapolation of the UT fit using the precisions expected at the next-generation
flavor facilities. Central values of all constraints are kept at the present averages.}
\label{fig:UTfuture}
\end{figure}

Next-generation flavor experiments are expected to increase the precision of
the determination of SM flavor parameters by one order of magnitude. This is
a first step in the quest for NP. Indeed, the high-correlated SM scenario, represented
by the overlapping of several CP-conserving and CP-violating constraints in 
the UT analysis, can easily break down in the presence of new flavor structures.
Figure~\ref{fig:UTfuture} illustrates a conceivable scenarios at the end of 
next-generation flavor experiments: some of the improved experimental
constraints of the UT no longer overlap signaling the presence of source of
flavor- and CP-violation beyond the CKM matrix. The generalized UT analysis,
exploiting the high-precision tree-level constraints, $\vert V_{ub}/V_{cb}\vert$ and
$\gamma$, can still determine the CKM parameters and, in addition, point out the
amplitudes deviating from the SM~\cite{Laplace:2002ik}.
%SM~\cite{Laplace:2002ik,Bona:2005eu,Bona:2006sa}.

On the other hand, it could happen that the improved UT experimental constraints
continue to overlap in one point. This would imply that NP is not showing up in
these constraints, mainly coming from mixing amplitudes, even with the increased accuracy.
Nonetheless the determination of the CKM parameters would still be improved resulting in
a better knowledge of the SM contribution to other FCNC and CP violating processes and
increasing in this way their sensitivity to NP.
For example, the error on the SM prediction of a golden mode for NP searches in
kaon physics, the CP-violating decay $K_L\to\pi^0\nu\bar\nu$, is presently dominated by
the uncertainty of the CKM parameters~\cite{Brod:2010hi}.

It is worth mentioning that the UT fit extrapolation shown above requires an improved
theoretical control of the hadronic uncertainty. Lattice QCD seems able to reach the required
accuracy on the time scale of future experiments~\cite{Bona:2007qt,O'Leary:2010af}.
Data-drive methods, based on the heavy quark expansion or on flavor symmetries,
are also promising.

% \subsection{Effective Field Theory Bounds On The New Physics Scale}
% The absence of NP signal in flavor data can be translated into a lower bound on
% the NP scale using an Effective Field Theory (EFT) approach. For example, 

\subsection{Supersymmetric Models}
The Minimal Supersymmetric Standard Model (MSSM) is the supersymmetric extension
of the SM with N=1 supersymmetry (SUSY), minimal particle content and R-parity conservation.
The particle spectrum is doubled by the inclusion of supersymmetric partners
and the Higgs sector is extended to include a second weak doublet (for a primer on the
MSSM see for instance ref.~\cite{hep-ph/9709356}). 
% The MSSM addresses many of the SM
% problems, providing a mechanism for a natural stabilization of the weak scale (provided
% new particles have masses at the TeV scale), a viable dark matter candidate, new sources of
% CP violation and allowing for the grand-unification of interactions.
While addressing most of the SM problems, the MSSM does not contain an explanation for
supersymmetry breaking and for flavor and CP violation. The two problems are intertwined,
as most of the new sources of flavor and CP violation are located in the soft SUSY breaking
terms. These terms are added to the MSSM Lagrangian to parameterized SUSY breaking in the
most general way which does not spoil the natural stabilization of the weak scale and is
compatible with gauge invariance and R-parity conservation.

The MSSM  introduces 105 new parameters in addition of 18 in the SM,
97 of which are related to flavor and CP violation: 21 squark and slepton masses,
36 new mixing angles and 40 new CP-violating phases measurable in sfermion
interactions~\cite{hep-ph/9709450}. Given this plethora of free parameters in the flavor
sector, it is not surprising that the MSSM suffers from the NP flavor problem, namely it
generates too large FCNC and CP violation for new particle masses at the TeV scale and
$\mathcal{O}(1)$ flavor parameters.

% A parameter counting shows that the MSSM  introduces
% 105 new parameters in addition of the 18 SM ones, 97 of which are related to flavor
% and CP violation: 21 squark and slepton masses, 36 new mixing angles and 40 new
% CP-violating phases measurable in sfermion interactions~\cite{hep-ph/9709450}.
% Given this plethora of free parameters in the flavor sector, it is not surprising that
% the MSSM suffers from the NP flavor problem, namely it generates too large FCNC and CP
% violation for new particle masses at the TeV scale and $\mathcal{O}(1)$ flavor parameters.

If SUSY is found by high-$p_T$ experiments at the LHC, establishing the flavor
structure of the MSSM is a major goal achievable at current and next-generation flavor
experiments with the added value of giving information on the SUSY breaking mechanism.
If, conversely, SUSY particles are too heavy to be produced at the LHC, there are
still chances to observe deviations from the SM in flavor physics.
\begin{figure}[h!]
\centerline{
\includegraphics[width=0.33\textwidth]{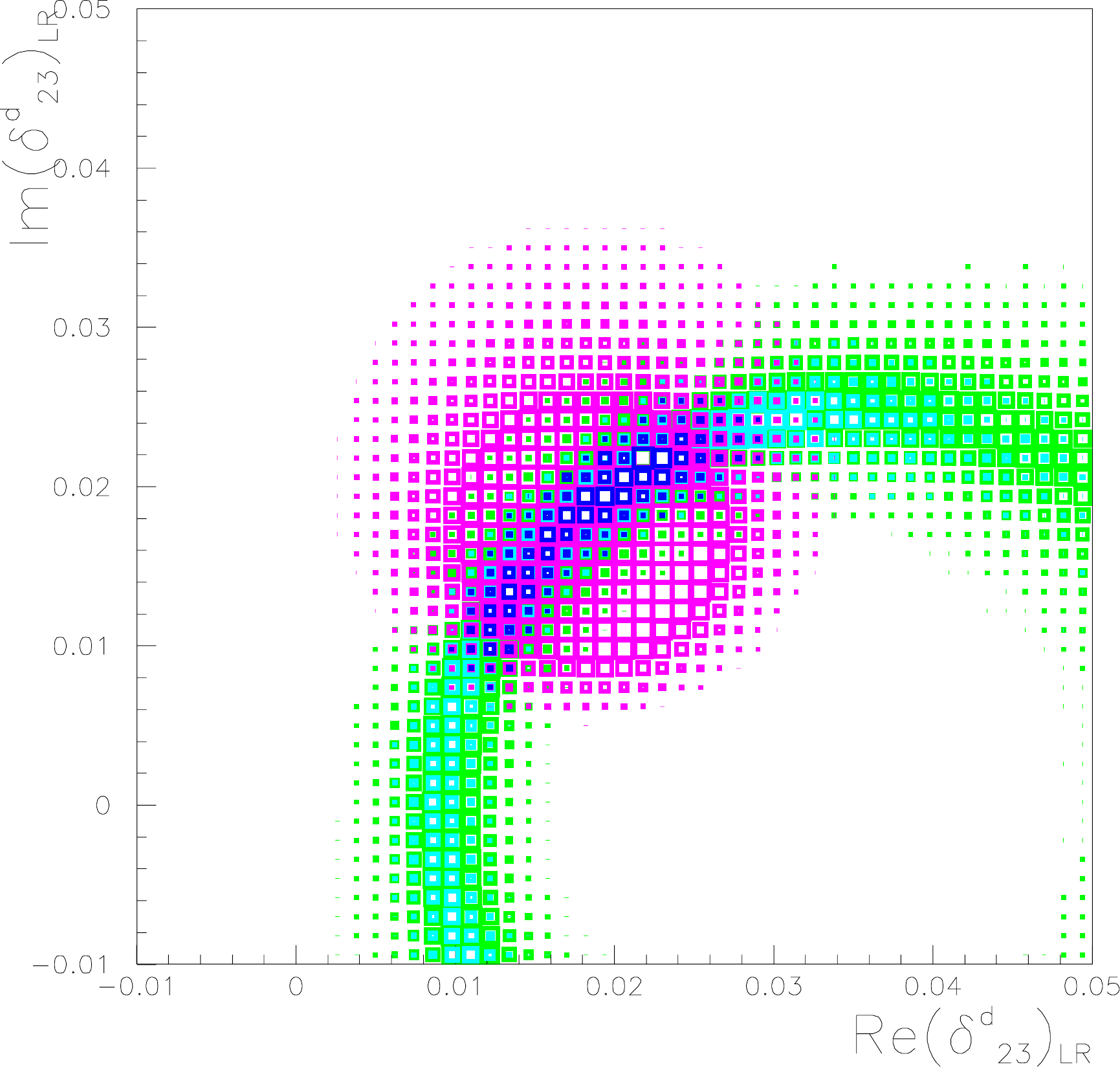}
\includegraphics[width=0.36\textwidth]{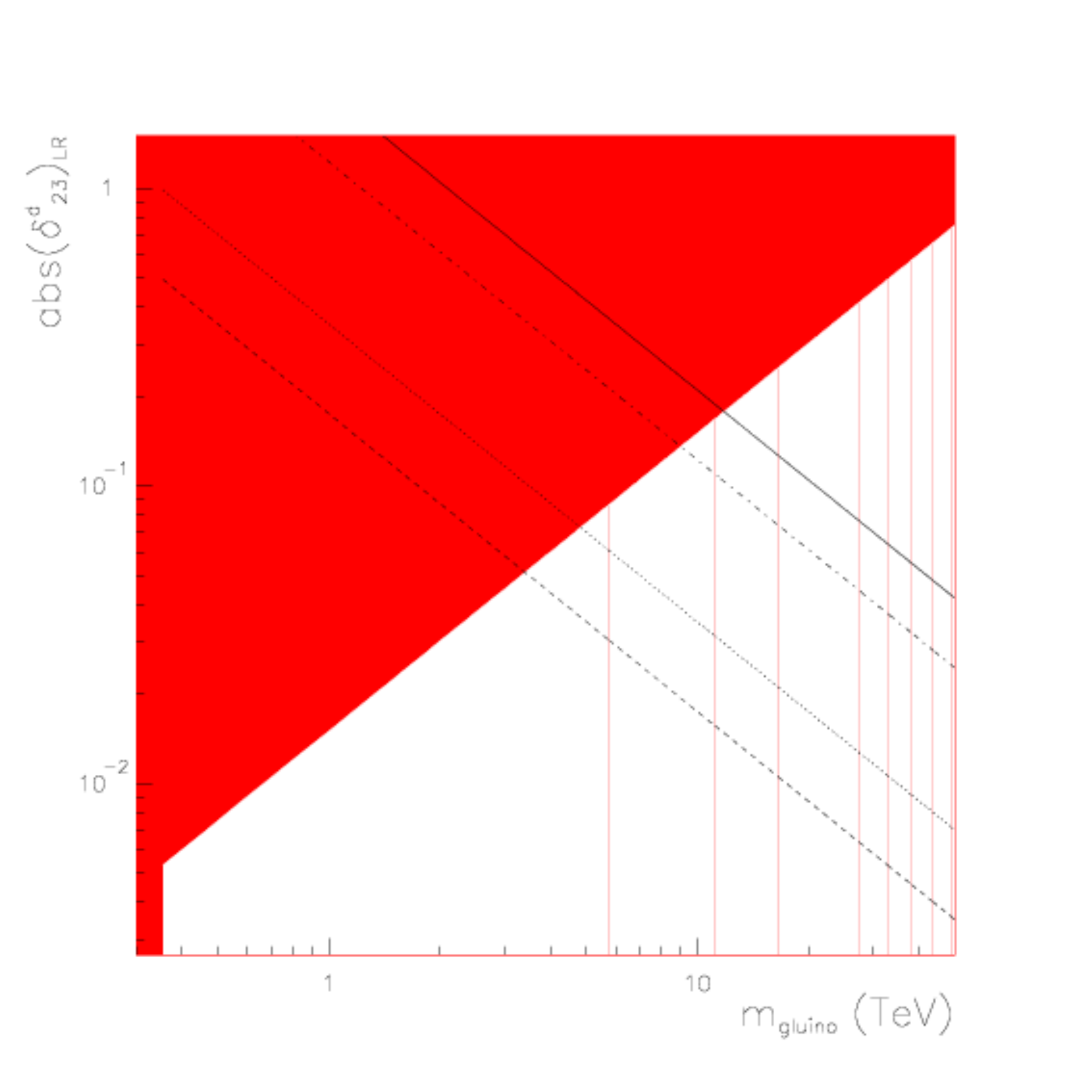}
}
\caption{Left plot: extraction of $(\delta_{23}^d)_{LR}$ from the measurements
of $a_\mathrm{CP}(B_d\to X_s\gamma)$ (magenta), $BR(B_d\to X_s\gamma)$ (green) and
$BR(B_d\to X_s\ell^+\ell^-)$ (cyan) with the errors expected at next-generation flavor
experiments. Central values are generated using $(\delta_{23}^d)_{LR}=0.028\,e^{i\pi/4}$
and squark and gluino masses at $1$ TeV. Right plot: region of the parameter space
where a non-vanishing $(\delta_{23}^d)_{LR}$ can be extracted with at least $3\sigma$
significance (in red).
}
\label{fig:d23lr}
\end{figure}
These two options are demonstrated in fig.~\ref{fig:d23lr} using the Mass Insertion
(MI) approximation~\cite{Hall:1985dx} to parameterize flavor-changing and
CP-violating effects in the squark sector and considering for simplicity only dominant
gluino contributions. The plot on the
left shows the reconstruction of the MI $(\delta_{23}^d)_{LR}$ from the measurements
of $A_\mathrm{CP}(B_d\to X_s\gamma)$, $BR(B_d\to X_s\gamma)$ and
$BR(B_d\to X_s\ell^+\ell^-)$ with the errors expected at next-generation
experiments. Central values are generated using $(\delta_{23}^d)_{LR}=0.028\,e^{i\pi/4}$
and squark and gluino masses at $1$ TeV. Both modulus and phase can be reconstructed 
with more than 5$\sigma$ significance. The plot on the right shows in red the region
where a non-vanishing absolute value of the MI can be extracted from the data with a
significance exceeding $3\sigma$ as a function of the gluino/squark mass. For masses
around $1$ TeV, one can measure MIs as small as $10^{-2}$. On the other hand, for larger
MIs, NP effects could be revealed for SUSY masses up to $10$ TeV.

\subsubsection{SUSY-breaking Scenarios}
SUSY-breaking models provide a complementary way to present the MSSM flavor phenomenology and
study the impact of next-generation experiments. Instead of a general parameterization
of the MSSM soft SUSY-breaking terms, one consider a SUSY-breaking mechanism, theoretically
or phenomenologically motivated, which reduces the number of independent parameters.
As most of the SUSY-breaking terms are also flavor-violating parameters, a correlation pattern
among flavor observables usually emerges, providing clues of the underlying theory.
As look-alikes are clearly possible, this way of studying the NP flavor structure and the
SUSY-breaking mechanism needs the measurement of several NP-sensitive observables.
In this respect, super flavor factories are ideal experimental facilities.

To illustrate this approach, we report the results of the study in
ref.~\cite{Altmannshofer:2009ne} summarized by table~\ref{tab:DNA}.
Several flavor observables are considered in a selection of SUSY models representative
of different scenarios. SUSY contributions to these observables are classified as
large, visible and small producing patterns which help identifying the model.

%%%%%%%%%%%%%%%%%%%%%%%%%%%%%%%%%%%%%%%%%%%%%%%%%%%%%%%%%%%%%%%%%%%%%%%%
\newcommand{\three}{{\color{red}$\bigstar\bigstar\bigstar$}}
\newcommand{\two}{{\color{blue}$\bigstar\bigstar$}}
\newcommand{\one}{{\color{black}$\bigstar$}}
%%%%%%%%%%%%%%%%%%%%%%%%%%%%%%%%%%%%%%%%%%%%%%%%%%%%%%%%%%%%%%%%%%%%%%%%
\begin{table}[h!]
\caption{
``DNA'' of flavor physics effects for the most interesting observables in a selection of SUSY
 models from ref.~\cite{Altmannshofer:2009ne}. \three\ signals large effects, \two\ visible but small
 effects and \one\ implies that the given model does not predict sizable effects in that
 observable.
\label{tab:DNA}}
%\addtolength{\arraycolsep}{4pt}
%\renewcommand{\arraystretch}{1.5}
\centering{\scriptsize 
\begin{tabular}{|l|c|c|c|c|c|}
\hline
&  AC & RVV2 & AKM  & $\delta$LL & FBMSSM 
\\
\hline
$D^0-\bar D^0$& \three & \one & \one & \one & \one \\
\hline
$ S_{\psi\phi}$ & \three & \three & \three & \one & \one  \\
\hline\hline
$S_{\phi K_S}$ & \three & \two & \one & \three & \three  \\
\hline
$A_{\rm CP}\left(B\rightarrow X_s\gamma\right)$ & \one & \one & \one & \three & \three \\
\hline
$A_{7,8}(B\to K^*\mu^+\mu^-)$ & \one & \one & \one & \three & \three \\
\hline
$A_{9}(B\to K^*\mu^+\mu^-)$ & \one & \one & \one & \one & \one \\
\hline
$B\to K^{(*)}\nu\bar\nu$  & \one & \one & \one & \one & \one \\
\hline
$B_s\rightarrow\mu^+\mu^-$ & \three & \three & \three & \three & \three \\
\hline
$\tau\rightarrow \mu\gamma$ & \three & \three & \one & \three & \three  \\
\hline
\end{tabular}
}
\end{table}
Another example is given in ref.~\cite{Aushev:2010bq} where a large number of correlations
between pairs of flavor observables is presented. 
% Indeed correlations of two or more
% observables are the most powerful tool to distinguish different models.

\subsubsection{Grand-Unified Models}
While in the SM the three gauge coupling constants do not exactly unify at any scale,
grand-unification can occur with SUSY. Typical SUSY-GUTs are based on
groups $SU(5)$ or $SO(10)$ breaking down to the SM group in one or more steps.
As far as flavor is concerned, the distinctive feature of grand-unified
models is that quark and leptons sit in the same representations of the gauge group
resulting in an interesting interplay between quark and lepton flavor violation, 
the details of which are model dependent.

% \begin{figure}[tb!]
% \centering\includegraphics[width=0.9\textwidth]{belleii.eps}
% \caption{Correlations between flavor observables in the case of the non-degenerate $SU(5)$
% SUSY GUT model with right-handed neutrinos from ref.~\cite{Aushev:2010bq}.
% Expected errors at SuperKEKB with an integrated luminosity of 50 ab$^{−1}$ are shown.}
% \label{fig:belleii}
% \end{figure}
% For example, among the many correlation plots in fig.~\ref{fig:belleii}, which refers to a
% grand-unified $SU(5)$ model with non-degenerate right-handed neutrinos~\cite{Aushev:2010bq},
% there are also non-trivial correlations between $\tau\to\mu$ and $b\to s$ transitions.
For example, many correlation plots are presented in ref.~\cite{Aushev:2010bq} for
few grand-unified $SU(5)$ models including non-trivial correlations between
$\tau\to\mu$ and $b\to s$ transitions.
\begin{figure}[tb!]
\centering\includegraphics[width=0.8\textwidth]{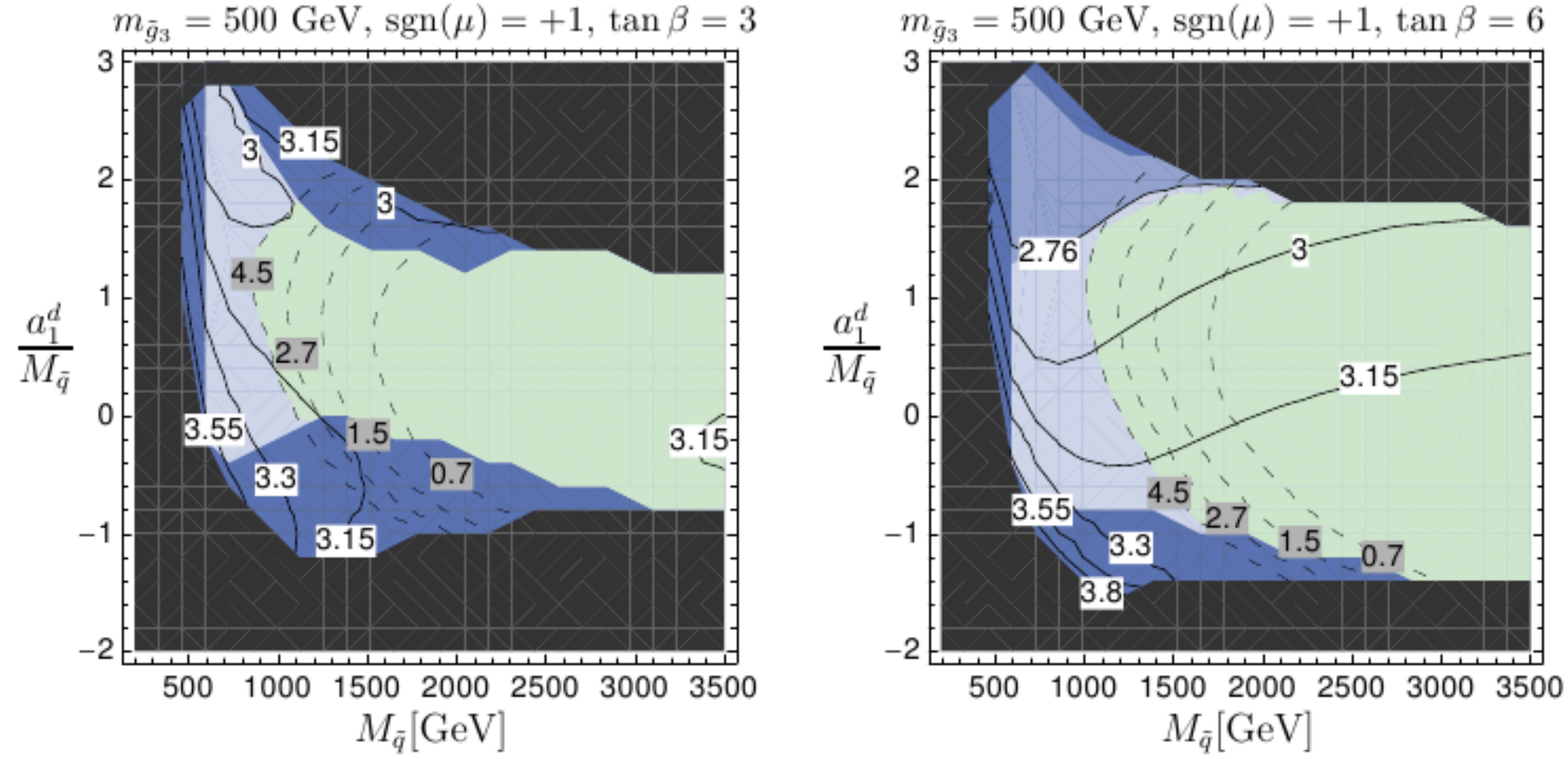}
\caption{Bounds on the parameter space of the $SO(10)$ SUSY model considered in
ref.~\cite{arXiv:1101.6047} for $m_{{\tilde g}_3}=500$ GeV and sgn$(\mu)=+1$ with
$\tan\beta=3$ (left) and $\tan\beta=6$ (right). $BR(B\to X_s\gamma)\times 10^4$ solid lines with white labels;
$BR(\tau\to\mu\gamma)\times 10^8$ dashed lines with gray labels. For more details, see ref.~\cite{arXiv:1101.6047}.}
\label{fig:so10}
\end{figure}

Similarly, quark and lepton flavor-changing transitions between the second and the
third generations put bounds on the
parameter space of the $SO(10)$ model considered in ref.~\cite{arXiv:1101.6047},
as shown in fig.~\ref{fig:so10}. In particular, the most effective constraint is
the branching ratio of $\tau\to\mu\gamma$, which will be measured at the super
flavor factories with a sensitivity of few in $10^{-9}$.

More generally, next generation flavor experiments, measuring both quark and lepton
flavor-changing processes, are a good testing ground for grand-unified SUSY
models.

\subsubsection{Minimal Flavor Violation}
Minimal Flavor Violation (MFV) provides a solution to the NP flavor
problem~\cite{Chivukula:1987py,D'Ambrosio:2002ex,Buras:2003jf}. The basic assumption of
MFV is that NP flavor
%problem~\cite{Chivukula:1987py,Hall:1990ac,Gabrielli:1994ff,Ciuchini:1998xy,D'Ambrosio:2002ex,Buras:2003jf}. The basic assumption of MFV is that NP flavor
effects are governed by the SM Yukawa couplings. MFV can be nicely formulated in the
effective field theory language: taking the Yukawa couplings as spurions of the SM flavor
group $G_\mathrm{flavor}$ in eq.~(\ref{eq:gflavor}), one can write
$G_\mathrm{flavor}$-invariant NP flavor-changing operators~\cite{D'Ambrosio:2002ex}.
These allow to prove that a NP scale at the TeV is compatible with present flavor
data in generic MFV models.
Popular SUSY models like mSUGRA  are MFV models. They are used as benchmarks for
LHC physics as they have a reduced parameter space. Yet, by construction, MFV models
produce small flavor effects (see for example ref.~\cite{hep-ph/0505110}) which could be
hard to reveal even at next-generation experiments, with $BR(B\to X_s\gamma)$ being the most
promising channel~\cite{Hitlin:2008gf}. There are however two cases where even MFV
models can give raise to large flavor effects: the large $\tan\beta$ regime 
in models with two Higgs doublets and the presence of non-negligible NP flavor-conserving
phases, which are not forbidden by the MFV condition. In the former case,
measurable deviations from the SM are expected to appear in $BR(B_s\to\mu^+\mu^-)$ and in
$BR(B_d\to\tau\nu)$. Indeed,  the large $\tan\beta$ scenario is already constrained by these
branching ratios~\cite{arXiv:0907.5135}
%ratios~\cite{arXiv:0907.5135,arXiv:0908.3470}
and greatly improved measurements are expected from LHCb and the super flavor factories.
In the latter case, flavor-diagonal phases can generate new CP-violating FCNC through
loop effects. For example, a sizable non-standard $A_\mathrm{CP}(B\to X_s\gamma)$
can be obtained in flavor-blind MSSM (FBMSSM) considered in ref.~\cite{Altmannshofer:2008hc}.
Clearly, as the sources of CP violation are flavor-conserving phases, the bounds coming
from Electric Dipole Moments (EDMs) are important and should be taken carefully
into account.
% 
% \begin{figure}[tb!]
% \centerline{
% \includegraphics[width=0.4\textwidth]{figures/ShK-ACP-A.eps}
% \includegraphics[width=0.4\textwidth]{figures/ShK-ACP-mu.eps}
% }
% \caption{Correlation between the mixing-induced CP asymmetry in $B\to\eta K_S$ and
% the direct CP asymmetry in $b\to s\gamma$ in the two scenarios with a complex $\mu$ term
% (left column) or complex $A_t$ term (right column) from ref.~\cite{arxiv:1102.0726}.
% The gray points are allowed by all constraints except the electric dipole moment of
% the electron, while the blue points are compatible with all constraints.
% See ref.~\cite{arxiv:1102.0726} for details.}
% \label{fig:acpvssetak}
% \end{figure}
Analogous results are obtained in the effective MFV model of ref.~\cite{arXiv:1102.0726},
which is based on a different flavor group, but produce a pattern of flavor-violating couplings
similar to MFV.

In summary, the main motivation for MFV is keeping the NP scale at the TeV implying that
new particles will be produced at the LHC. In such a case, the next-generation
flavor experiments will have the task to study their flavor properties, trying to
elucidate the pattern of flavor suppression. In this respect, establishing the MFV nature
of NP could be a very hard task, involving small and correlated NP effects in $K$, $B_d$ and
$B_s$ physics. On the other hand, even MFV models could produce large flavor effects
is special regimes and super flavor factories will be able to perform key
precision measurements to reveal them.

\begin{figure}[h!]
\centering\includegraphics[width=.5\textwidth]{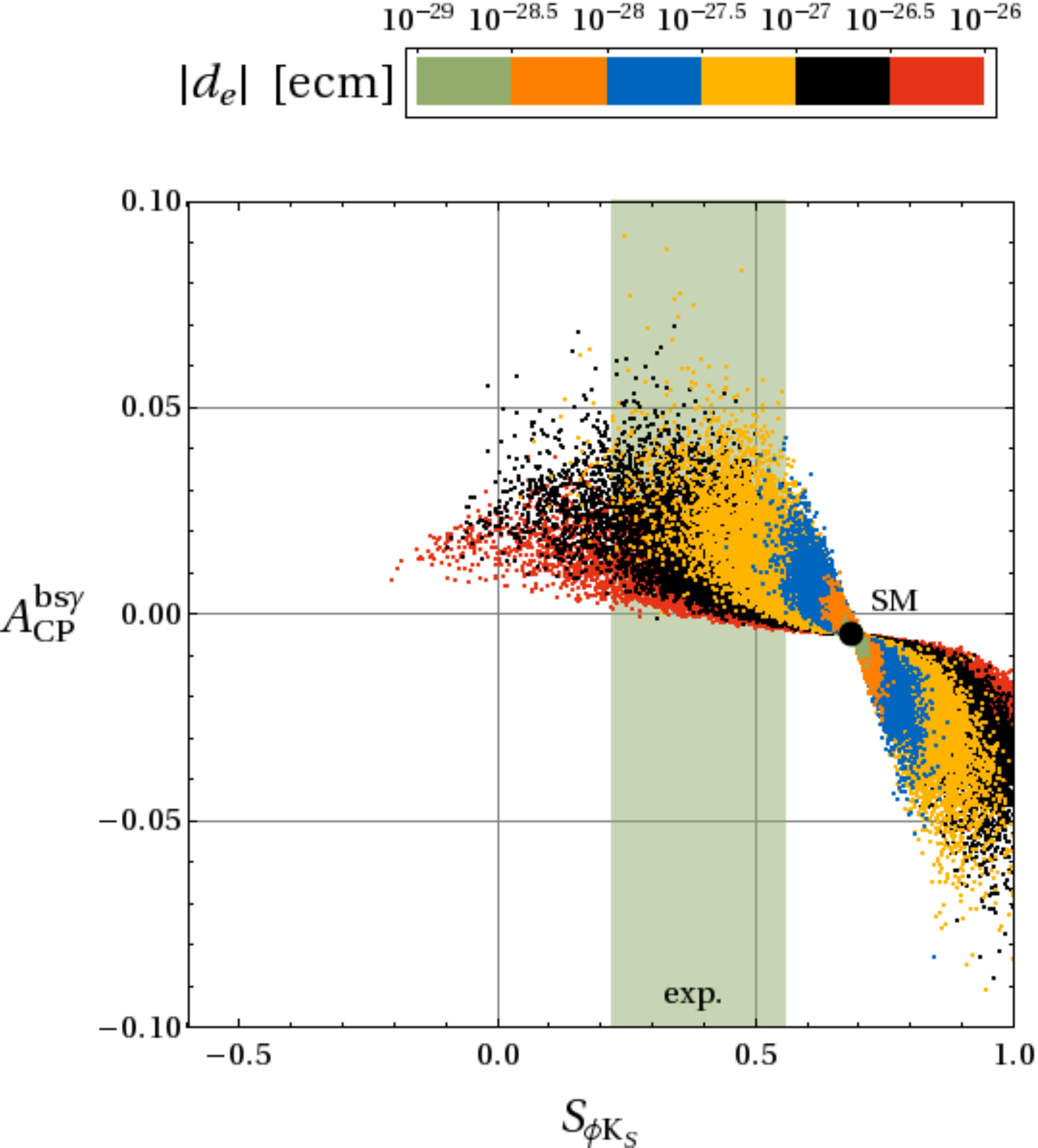}
\caption{Correlation between the CP asymmetries $A_\mathrm{CP}(B\to X_s\gamma)$ and
$S_{B_d\to\phi K_S}$ in the FBMSSM \cite{Altmannshofer:2008hc}. The various colors indicate
the predicted lower bound on the electron EDM $|d_e|$. \label{fig:FBMSSM}}
\end{figure}

\subsection{Standard Model With Four Generations}
\begin{figure}[h!]
\begin{center}
\includegraphics[width=.45\textwidth]{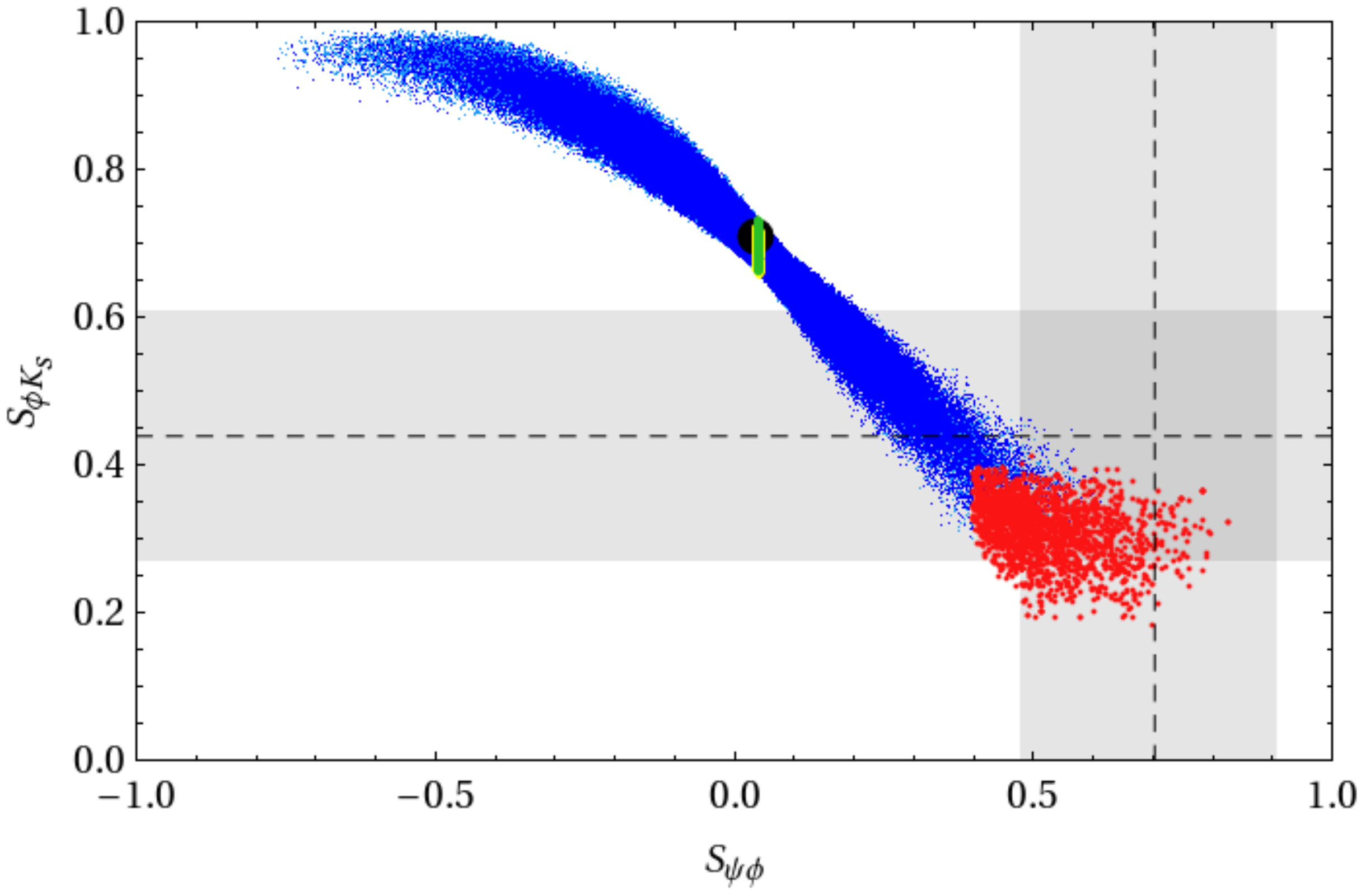}
\includegraphics[width=.45\textwidth]{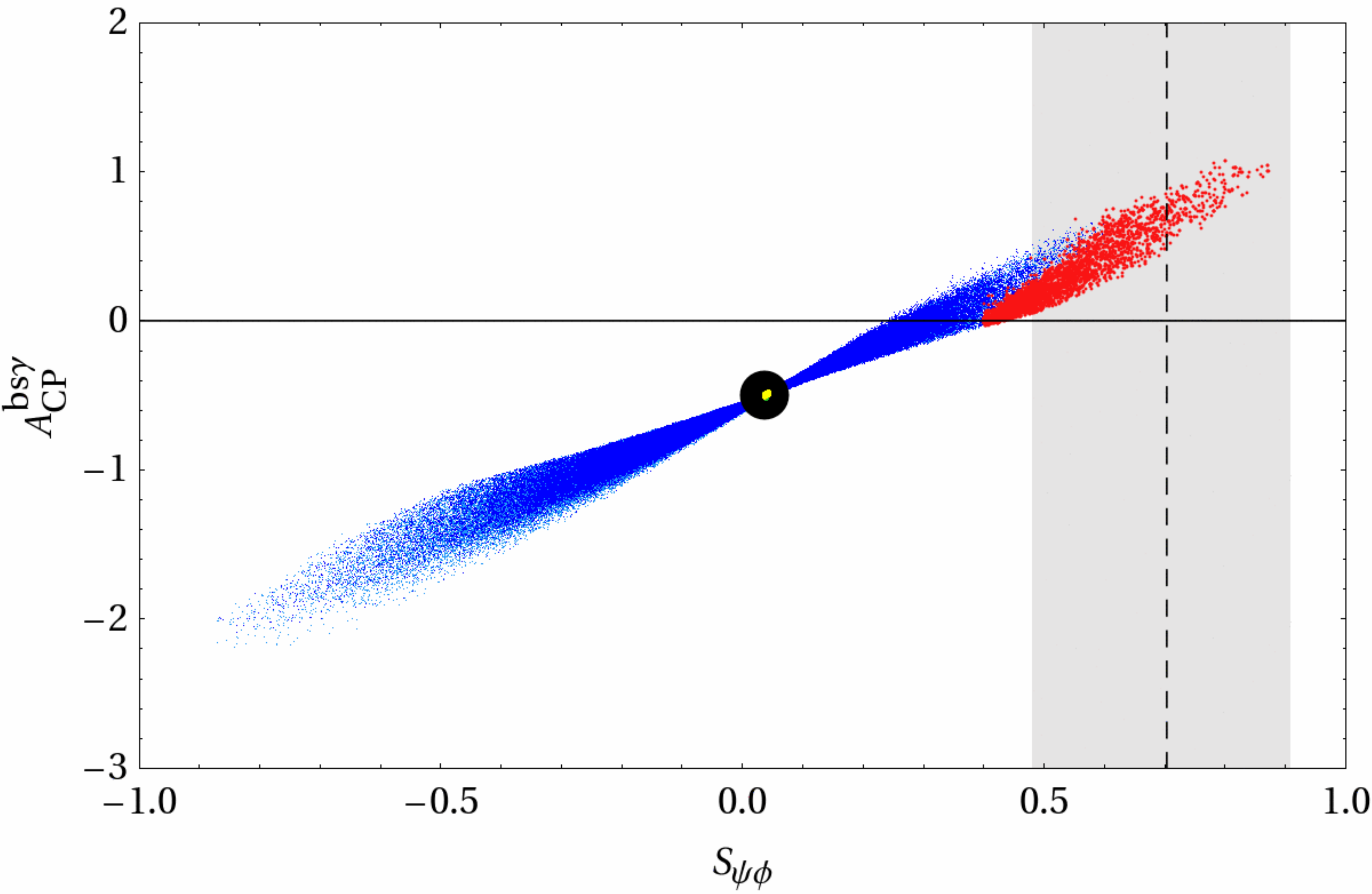}
\caption{Correlation of  $S_{\psi\phi}$ with $S_{\phi K_S}$ (left) and 
 $a_\mathrm{CP}(b\to s\gamma)$ (right) in the SM4 from ref.~\cite{Buras:2010pi}.
\label{fig:SM4}}
\end{center}
\end{figure}

The interest in SM4, namely the extension of the SM which includes a fourth generation of heavy quarks
and leptons, was revived in recent years~\cite{Hou:2005yb,Soni:2010xh,Buras:2010pi}.
%years~\cite{Hou:2005yb,Hou:2005yb,Hou:2006mx,Herrera:2008yf,Soni:2010xh,Buras:2010pi}.
While leaving several SM problems unresolved, SM4 has the some interesting
features: it allows for a heavier Higgs boson, relaxing the indication of a light Higgs coming from
the fit of electroweak precision data, and can accommodate a large CP violation in $B_s$ mixing, as
hinted by recent Tevatron data. In addition, the new fermions can be produced and detected at the LHC,
so that SM4 will be found or excluded soon. In the former case, besides the direct observation,
measurable NP flavor effects are expected in time-dependent angular analysis of $B_s\to J/\psi\phi$, precisely
measured at LHCb. Corresponding deviations from the SM in $b\to s$ transitions involving the $B_d$ meson
can be measured at the super flavor-factories, as shown in fig.~\ref{fig:SM4}.
\vspace*{0,5CM}

\subsection{Little Higgs Models}
Little Higgs models assume that the Higgs field is a pseudo-Goldstone
boson associated to the breaking of a global symmetry~\cite{hep-th/0104005}.
The careful choice of the breaking mechanism can naturally stabilize the electroweak scale,
raising the cutoff of the theory up to $10$ TeV region. 
While several variants of little Higgs models exist, the Littlest Higgs model with
T-parity (LHT) has the additional features of fulfilling the constraints
coming from electroweak precision data for a relatively small symmetry breaking scale
$f$ and providing a dark matter candidate. In this model, the exchange of new heavy quark,
scalars and gauge bosons gives rise to new flavor effects.
FCNC and CP violation in the LHT have been extensively studied in
refs~\cite{Blanke:2006sb,Blanke:2009am}: large effects are possible in rare kaon decays, such
as $K_L\to\pi^0\nu\bar\nu$ and $K^\to\pi^+\nu\bar\nu$ and $B_s$--$\bar B_s$ mixing,
while $B_d$ processes typically receive small corrections.
\begin{table}[h!]
\caption{Comparison of various ratios of branching ratios in the LHT model
($f=1$ TeV) \cite{Blanke:2009am} and in the MSSM without \cite{Ellis:2002fe} and
with \cite{Paradisi:2005tk} significant Higgs contributions.\label{tab:ratios}}
%{\renewcommand{\arraystretch}{1.5}
\begin{center}
\begin{tabular}{|c|c|c|c|}
\hline
\multirow{2}{2cm}{\centering ratio} & \multirow{2}{2cm}{\centering LHT}  & MSSM     & MSSM \\
      &      & (dipole) & (Higgs)\\\hline\hline
$\frac{BR(\tau^-\to e^-e^+e^-)}{BR(\tau\to e\gamma)}$   & 0.04\dots0.4     &
$\sim1\cdot10^{-2}$ & ${\sim1\cdot10^{-2}}$\\
$\frac{BR(\tau^-\to \mu^-\mu^+\mu^-)}{BR(\tau\to \mu\gamma)}$  &0.04\dots0.4     &
$\sim2\cdot10^{-3}$ & $0.06\dots0.1$ \\\hline
$\frac{BR(\tau^-\to e^-\mu^+\mu^-)}{BR(\tau\to e\gamma)}$  & 0.04\dots0.3     &
$\sim2\cdot10^{-3}$ & $0.02\dots0.04$ \\
$\frac{BR(\tau^-\to \mu^-e^+e^-)}{BR(\tau\to \mu\gamma)}$  & 0.04\dots0.3    &
$\sim1\cdot10^{-2}$ & ${\sim1\cdot10^{-2}}$\\
$\frac{BR(\tau^-\to e^-e^+e^-)}{BR(\tau^-\to e^-\mu^+\mu^-)}$     & 0.8\dots2.0   &
$\sim5$ & 0.3\dots0.5\\
$\frac{BR(\tau^-\to \mu^-\mu^+\mu^-)}{BR(\tau^-\to \mu^-e^+e^-)}$   & 0.7\dots1.6    &
$\sim0.2$ & 5\dots10 \\\hline
\end{tabular}
\end{center}%\renewcommand{\arraystretch}{1.0}
%}
\end{table}

Lepton flavor violation in $\tau$ decays is particularly interesting in LHT. Contrary
to the SM and many of its extensions including the MSSM, the ratio of 
$BR(\tau\to 3\mu)$ over $BR(\tau\to\mu\gamma)$ is not governed by the electromagnetic
constant $\alpha_e$, see table~\ref{tab:ratios}.
The rate of $\tau\to\mu\gamma$ can be large enough to be measured at super flavor
factories if the symmetry breaking scale $f$ is around $500$ GeV. For larger scales,
$BR(\tau\to\mu\gamma)$ becomes too small, but $BR(\tau\to 3\mu)$ could still be
measurable. In any case the large ratio $BR(\tau\to 3\mu)/BR(\tau\to\mu\gamma)$ is a
distinctive feature which could tell the LHT apart from other extensions of the SM.

\begin{table}[h!]
\caption{Maximal values on LFV $\tau$ decay branching ratios in the LHT model, for two
different values of the scale $f$, after imposing the constraints on $\mu\to e\gamma$ and
$\mu^-\to e^-e^+e^-$ from ref.~\cite{Blanke:2009am}.  \label{tab:bounds}}
\begin{center}
{\small
\begin{tabular}{|c|c|c|c|}
\hline
\multirow{2}{2cm}{\centering decay} & \multirow{2}{2,5cm}{\centering $f=1000$ GeV} & \multirow{2}{2,5cm}{\centering $f=500$ GeV} & Super$B$ \\
 & & & sensitivity \\\hline
$\tau\to e\gamma$ & $8\cdot 10^{-10}$  & ${2\cdot 10^{-8}}$ & ${2\cdot10^{-9}}$  \\
$\tau\to \mu\gamma$ & $8\cdot 10^{-10}$  &$2\cdot 10^{-8}$   &${2\cdot10^{-9}}$ \\
$\tau^-\to e^-e^+e^-$ & $1\cdot10^{-10}$  & ${2\cdot10^{-8}}$   & $2\cdot10^{-10}$ \\
$\tau^-\to \mu^-\mu^+\mu^-$ & $1\cdot10^{-10}$  & ${2\cdot10^{-8}}$   & $2\cdot10^{-10}$ \\
%$\tau^-\to e^-\mu^+\mu^-$ & $1\cdot10^{-10}$ & ${2\cdot10^{-8}}$   & \\
%$\tau^-\to \mu^-e^+e^-$ & $1\cdot10^{-10}$ & ${2\cdot10^{-8}}$  & \\
%$\tau^-\to \mu^-e^+\mu^-$ & $6\cdot10^{-14}$ & ${1\cdot10^{-13}}$ & \\
%$\tau^-\to e^-\mu^+e^-$ & $6\cdot10^{-14}$ &${1\cdot10^{-13}}$   &  \\
%$\tau\to\mu\pi$ & $4\cdot10^{-10} $  & ${5\cdot10^{-8}}$  & \\
%$\tau\to e\pi$ & $4\cdot10^{-10} $ & ${5\cdot10^{-8}}$   & \\
$\tau\to\mu\eta$ & $2\cdot10^{-10}$  & ${2\cdot10^{-8}}$  & $4\cdot10^{-10}$ \\
$\tau\to e\eta$ & $2\cdot10^{-10}$  & ${2\cdot10^{-8}}$  & $6\cdot10^{-10}$ \\
%$\tau\to \mu\eta'$ & $1\cdot10^{-10}$ & ${2\cdot10^{-8}}$  & \\
%$\tau\to e\eta'$ & $1\cdot10^{-10}$ & ${2\cdot10^{-8}}$   & \\
\hline
\end{tabular}
}
\end{center}
\end{table}

LHT effects in $D$ physics have been studied in ref.~\cite{Bigi:2009df}. A large CP violation,
a clear NP signal, can be present both in the mixing amplitude ($A_\mathrm{SL}$) and in the
time-dependent CP asymmetry of $D\to K_S\phi$ ($S_{K_S \phi}$).
%, as shown in fig.~\ref{fig:LHT-D}. The plotted correlation assumes that there is no direct CP violation in the $D \to K_S \phi$ decay.
The possibility to measure $D$ time-dependent CP asymmetries is under study at SuperB.
% \begin{figure}[tb!]
% \begin{center}
% \includegraphics[width=.7\textwidth]{aSL-SDKphi}
% \caption{Correlation between the CP asymmetries $a_\mathrm{SL}$ and $S_{K_S \phi}$ in the
% LHT model from ref.~\cite{Bigi:2009df}.\label{fig:LHT-D}}
% \end{center}
% \end{figure}

\subsection{Warped Extra Dimensions}
Randall-Sundrum models of extra dimensions exploits the geometry of the fifth dimension to
naturally stabilize the Fermi scale~\cite{Randall:1999ee}. In addition, by putting fermions in the bulk,
one can address the SM flavor problem: the hierarchies in quark masses and mixings are obtained
from the localization of fermions along the fifth dimension~\cite{Gherghetta:2000qt}.

\begin{figure}[h!]
\begin{center}
\centering\includegraphics[width=.5\textwidth]{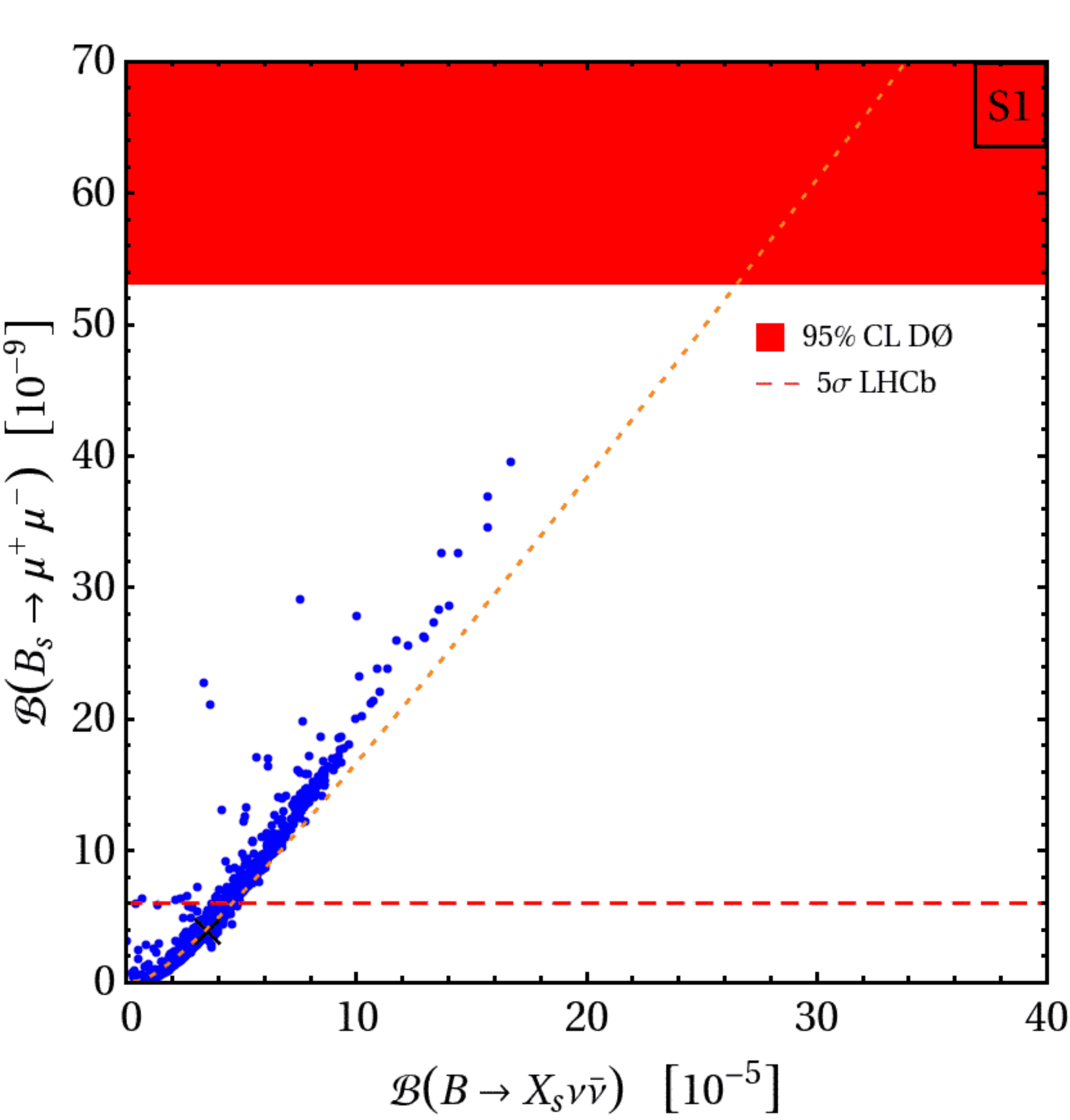}
\caption{Correlation between the branching ratios for $B_s\to \mu^+\mu^-$ and
$B\to X_s\nu\bar\nu$ in the minimal RS model from ref.~\cite{Bauer:2009cf}.\label{fig:minRS}}
\end{center}
\end{figure}

New flavor effects in these models arise from the exchange of Kaluza-Klein excitations.
They depend crucially on the specific realization of the model, although in
general they are severely constrained by CP violation in $K$--$\bar K$ mixing.
In the minimal scenario with only the SM gauge group in the bulk, one finds that large effects
are possible in both $B$ and $K$ decays~\cite{Agashe:2004ay,Blanke:2008zb,Bauer:2009cf}, see
for example the correlation between $BR(B_s\to\mu^+\mu^-)$ and $BR(B\to X_s\nu\bar\nu)$
in fig.~\ref{fig:minRS}. The former $BR$ will be measured soon at LHCb, while the transition
$b\to s\nu\bar\nu$ can be measured in the exclusive decays $B\to K^{*}\nu\bar\nu$ with full
statistics at SuperB.

Other realizations, such as the custodially-extended bulk gauge symmetry, suppress the NP
contributions to $B$ FCNCs, although large effects in kaons and in $B_s$--$\bar B_s$ mixing
are still possible~\cite{Blanke:2008zb}.

\section{Conclusions}
\label{sec:conclusions}
In this review we have considered a large subset of flavor measurements,
briefly discussing their present status and the expected sensitivities at
next-generation experiments, including possible systematic or theoretical limitations. 
We have then presented some examples of flavor effects generated in a selection of NP models
which could be measurable in the future. In particular, we selected those
correlations among observables which could provide a distinctive mark of the NP model,
showing how flavor physics could help telling it apart from the others.

The next decade will be crucial for the future of particle physics in general
and of collider physics in particular. The LHC already
started looking for the Higgs boson and for new particles at the TeV scale.
In this context, flavor physics could have a prominent role in characterizing
a TeV NP and looking for signals from the multi-TeV region.
It is even possible that the first evidence for NP come from flavor, with LHCb
measuring NP-sensitive channels like $B_s\to\mu\mu$ and $B_s\to J/\psi\phi$.
In any case, reconstructing the NP Lagrangian from the data is a collective
effort requiring the interplay of both direct and indirect searches, together
with an accurate theoretical control of the SM contributions.
Improved experimental results at the energy and intensity frontiers and
the most advanced theoretical techniques are therefore required ingredients
to go steadily beyond the SM. 

Precision flavor physics can certainly give a major contribution to
this effort. Next-generation experiments will offer new opportunities by pushing
existing flavor measurements to an unprecedented accuracy and by opening the
possibility to measure new NP-sensitive observables in all flavor sectors.

\end{document}